\documentclass[useAMS,usenatbib]{emulateapj}
\usepackage{comment}
\usepackage{xspace}

\newcommand{\sgra}{Sgr A$^*$\xspace}

\shorttitle{X-ray Flares from \sgra}
\shortauthors{Neilsen et al.}

\begin{document}

\title{A \textit{Chandra}/HETGS Census of X-ray Variability From \sgra During 2012}

\author{J.\ Neilsen\altaffilmark{1,2}, M.~A.\ Nowak\altaffilmark{2}, C.\ Gammie\altaffilmark{3,4}, J.\ Dexter\altaffilmark{6}, S.\ Markoff\altaffilmark{7}, D.~Haggard\altaffilmark{8},  S.\ Nayakshin\altaffilmark{5}, Q.~D.\ Wang\altaffilmark{10}, N.\ Grosso\altaffilmark{11}, D.\ Porquet\altaffilmark{11}, J.~A.\ Tomsick\altaffilmark{12}, N. Degenaar\altaffilmark{13}, P.~C.\ Fragile\altaffilmark{14}, R. Wijnands\altaffilmark{7}, J.~M.\ Miller\altaffilmark{13}, F.~K.\ Baganoff\altaffilmark{2}}
\altaffiltext{1}{Einstein Fellow, Boston University Department of Astronomy, Boston, MA 02215, USA; neilsenj@bu.edu}
\altaffiltext{2}{MIT Kavli Institute for Astrophysics and Space Research, Cambridge, MA 02139, USA; mnowak@space.mit.edu, houck@space.mit.edu}
\altaffiltext{3}{Department of Astronomy, University of Illinois Urbana-Champaign, 1002 W. Green St., Urbana, IL 61801, USA; gammie@illinois.edu}
\altaffiltext{4}{Department of Physics, University of Illinois Urbana-Champaign 1110 W. Green St., Urbana, IL 61801, USA}
\altaffiltext{5}{Department of Physics \& Astronomy, University of Leicester, Leicester LE1 7RH; sn85@leicester.ac.uk}
\altaffiltext{6}{Theoretical Astrophysics Center and Department of Astronomy, University of California, Berkeley, CA 94720-3411, USA}
\altaffiltext{7}{Astronomical Institute, ``Anton Pannekoek'', University of Amsterdam, Postbus 94249, 1090 GE Amsterdam, The Netherlands; s.b.markoff@uva.nl, r.a.d.wijnands@uva.nl}
\altaffiltext{8}{Center for Interdisciplinary Exploration and Research in Astrophysics, Physics and 
Astronomy Department, Northwestern University, Evanston, IL 60208, USA; CIERA Postdoctoral Fellow}  
%\altaffiltext{9}{Shanghai Astronomical Observatory, CAS; Shanghai 200030, China; fyuan@shao.ac.cn}
 \altaffiltext{10}{Department of Astronomy, University of Massachusetts, Amherst, MA 01002, USA; wqd@astro.umass.edu}
\altaffiltext{11}{Observatoire Astronomique de Strasbourg, Universit\'e de Strasbourg, CNRS, UMR 7550, 11 rue de l'Universit\'e, 67000 Strasbourg, France; nicolas.grosso@astro.unistra.fr, delphine.porquet@astro.unistra.fr}
\altaffiltext{12}{Space Sciences Laboratory, 7 Gauss Way, University of California, Berkeley, CA 94720-7450; jtomsick@ssl.berkeley.edu}
\altaffiltext{13}{Hubble Fellow, Department of Astronomy, University of Michigan, 500 Church Street, Ann Arbor, MI 48109, USA; degenaar@umich.edu, jonmm@umich.ed}
\altaffiltext{14}{Department of Physics \& Astronomy, College of Charleston, Charleston, SC 29424, USA; fragilep@cofc.edu}

\begin{abstract}
We present the first systematic analysis of the X-ray variability of \sgra during the \textit{Chandra X-ray Observatory}'s 2012 \sgra X-ray Visionary Project (XVP). With 38 High Energy Transmission Grating Spectrometer  (HETGS) observations spaced an average of 7 days apart, this unprecedented campaign enables detailed study of the X-ray emission from this supermassive black hole at high spatial, spectral and timing resolution. In 3 Ms of observations, we detect 39 X-ray flares from \sgra, lasting from a few hundred seconds to approximately 8 ks, and ranging in $2-10$ keV luminosity from $\sim10^{34}$ erg s$^{-1}$ to $2\times10^{35}$ erg s$^{-1}.$ Despite tentative evidence for a gap in the distribution of flare peak count rates, there is no evidence for X-ray color differences between faint and bright flares. Our preliminary X-ray flare luminosity distribution $dN/dL$ is consistent with a power law with index $-1.9^{+0.3}_{-0.4}$; this is similar to some estimates of \sgra's NIR flux distribution. The observed flares contribute one-third of the total X-ray output of \sgra during the campaign, and as much as 10\% of the quiescent X-ray emission could be comprised of weak, undetected flares, which may also contribute high-frequency variability. We argue that flares may be the only source of X-ray emission from the inner accretion flow.
\end{abstract}
                 

\section{INTRODUCTION}\label{sec:intro}

Since the launch of the \textit{Chandra X-ray Observatory} and \textit{XMM-Newton} in 1999, X-ray observations of \sgra, the $4\times10^{6}M_{\odot}$ black hole at the center of our Galaxy, have revealed a supermassive black hole deep in quiescence, a profound inactivity punctuated roughly once a day by rapid flares (e.g.\ \citealt{Baganoff01,Melia01,Baganoff03}; \citealt*{Genzel10,Markoff10} and references therein). But rather than clarifying the nature of the X-ray emission, the flares from \sgra have added to the puzzle of its extremely low luminosity. 

In weakly accreting black hole systems at all mass scales, there is a well-known three-way correlation between black hole mass, X-ray luminosity, and radio luminosity (the fundamental plane of black hole activity; \citealt*{Merloni03,Falcke04,Kording06,Gultekin09}; see also \citealt{Gallo06}; \citealt*{Yuan09}). However, as demonstrated by \citet{Markoff05b}; \citet*{Kording06}; \citet{Plotkin12}, \sgra only lies on (or near) this fundamental plane during its X-ray flares; in quiescence, the supermassive black hole is a notable outlier. Thus it is evident that studies of the quiescent emission and the flare emission from \sgra probe inherently different physical processes, and that through the flares, we can explore the connections between \sgra and weakly accreting black holes, such as those in low-luminosity AGN (LLAGN, e.g.\ \citealt*{Ho08,Markoff10,Yuan11} and references therein). The rarity of strong flares in other LLAGN (such as M81: \citealt{Markoff08}; M31: \citealt{Garcia10}, although see \citealt{Li11}; see also \citealt*{Yuan04}) places \sgra in a unique position (apparently unable to produce continuous X-ray emission in accordance with the fundamental plane) and highlights the importance of understanding the physics of flares.

One such flare (\citealt{Nowak12}) began on 2012 February 9 around 14:28:19 (UTC). With \textit{Chandra} looking on, the $2-10$ keV X-ray luminosity rose by a factor of at least $\sim130$ over the course of an hour, reaching a value of about $5\times10^{35}$ erg s$^{-1}$ for a distance of 8 kpc. At its peak, the X-ray emission from this flare (the brightest X-ray flare ever observed from \sgra) equaled the normal bolometric luminosity of this supermassive black hole. Within another 30 minutes, it had vanished. But beyond the strong variability on short time scales, what makes this flare and its record-breaking intensity so remarkable is that it topped out at a maximum luminosity of $10^{-9}L_{\rm Edd}$ (still far below the luminosities originally studied on the fundamental plane; see also \citealt{Porquet03}). The majority of X-ray flares, however, are much fainter, with average luminosities around $10\times$ the quiescent level. 

A summary of the phenomenology of flares from \sgra is provided by \citet{Dodds-Eden09}. Studies of the Galactic Center detect approximately four times more flares in the near-infrared (NIR) band than in X-rays (e.g.\ \citealt{Genzel03,Eckart06}); the evidence suggests that every X-ray flare is accompanied by a more-or-less simultaneous NIR flare, while some NIR flares have no X-ray counterpart (e.g.\ \citealt{Hornstein07}). In addition, while the spectral indices in these bands are similar, the X-ray flux as seen by \textit{Chandra}, \textit{XMM-Newton}, and \textit{Swift} is not necessarily consistent with the extrapolated NIR spectrum (e.g.\ \citealt{Eckart06}). In both the infrared (e.g.\ \citealt{Hornstein07,Bremer11}) and the X-ray (\citealt{Porquet03,Porquet08,Nowak12,Degenaar12_arxiv}), there has been some debate regarding the dependence of the flares' spectral indices on luminosity. Given the complex relationship between the NIR and X-ray emission, these dependences may be key to understanding flare physics.

Despite the focused attention on both the quiescent X-ray emission and the flares from \sgra during the last decade, neither is completely understood. For reference, the quiescent emission includes all X-ray emission within $1.25\arcsec\approx 6400$ AU $\approx 1.6 \times 10^5 R_g$ of \sgra, where $R_g= G M_{\rm BH}/c^2$ is the gravitational radius and $M_{\rm BH}$ is the black hole mass. This includes not only any emission from the hot inner accretion flow, but also from the rest of the accretion flow on scales approaching the Bondi radius (as well as stars, compact objects, and extended diffuse emission integrated along the line of sight, but see Section \ref{sec:flarepsf} and \citealt{Shcherbakov10}). The quiescent X-ray spectrum can be fitted as thermal bremsstrahlung ($kT=2-3.5$ keV), and has shown so little variability over the last 14 years of monitoring that the only plausible physical scenarios for its origin are thermal plasma from large scales or a cluster of coronally active stars (e.g.\ \citealt{Falcke00,Quataert02,Baganoff03}; \citealt*{Yuan02,Yuan03,Liu04}; \citealt{Xu06}; \citealt*{Sazonov12}; \citealt{Nowak12}; but see \citealt{Wang13}, in press).

A rich diversity of models has been proposed to explain the flares, both in terms of their energy injection mechanisms, ranging from magnetic reconnection, stochastic acceleration, or shocks in the accretion flow or jet to tidal disruption of asteroids, and in terms of their radiation mechanisms, for which the X-ray photon index $\Gamma\sim2$ could be due to direct synchrotron, synchrotron self-Compton (SSC), or inverse Compton processes (e.g.\ \citealt{Markoff01,Liu02,Liu04,Yuan03,Yuan04,Eckart04,Eckart06,Marrone08}; \citealt*{Cadez08}; \citealt{Kostic09,Dodds-Eden09,Yuan09b}; \citealt*{Zubovas12}; \citealt{Witzel12,Yusef-Zadeh12,Nowak12} and references therein). The reduced rate of X-ray flares relative to the NIR may be explained if the high-energy emission is dominated by SSC, such that only events for which the column density of synchrotron-emitting electrons is sufficiently large will produce X-ray flares observable above the quiescent emission (e.g.\ \citealt{Marrone08} and references therein).

In order to address both the origin of the quiescent emission and the physics of the flares (as well as to continue detailed study of diffuse emission and transients in the Galactic Center; e.g. \citealt{Baganoff03,Muno04,Park04,Muno07,Muno09,Ponti10,Capelli12}), \textit{Chandra} undertook an unprecedented X-ray Visionary Project (XVP) in 2012 to observe \sgra for 3 Ms at high spectral resolution with the High Energy Transmission Grating Spectrometer (HETGS; \citealt{C05}). The excellent X-ray spectral resolution provides a robust probe of the nature of the accretion flow (\citealt{Xu06,Young07,Sazonov12}), and the high cadence and coverage of observations and the long exposure time are ideal for the first-ever systematic study of X-ray variability and flares from \sgra.

This paper is the first systematic study of X-ray flares from \sgra during the 2012 \textit{Chandra} XVP. Here we focus on quantifying the properties (i.e.\ intensity and duration) of the flares observed during our 38 \textit{Chandra} observations and characterizing the relationship between the quiescent emission and the X-ray flares. Although prior \textit{Chandra}, \textit{XMM-Newton}, and \textit{Swift} observations have detected many flares, they used different detectors and have very different spectral responses and levels of pileup (Section \ref{sec:pile}); we leave a comprehensive analysis for future work. The paper is organized as follows. In Sections \ref{sec:obs} and \ref{sec:lc} we describe the \textit{Chandra} observations and data reduction and our method for identifying flares. In Section \ref{sec:stat}, we assess the statistical properties of the flares, including their peak brightnesses, fluences, durations, estimated luminosities, and X-ray colors. We also explore the variability caused by undetected flares. In Section \ref{sec:discuss}, we discuss our results in the context of flare models, multiwavelength flux distributions, and the contribution of weak flares to the quiescent X-ray emission.

\section{OBSERVATIONS AND DATA REDUCTION}
\label{sec:obs}

\setcounter{footnote}{0}
\begin{figure*}
\centerline{\includegraphics[width=\textwidth]{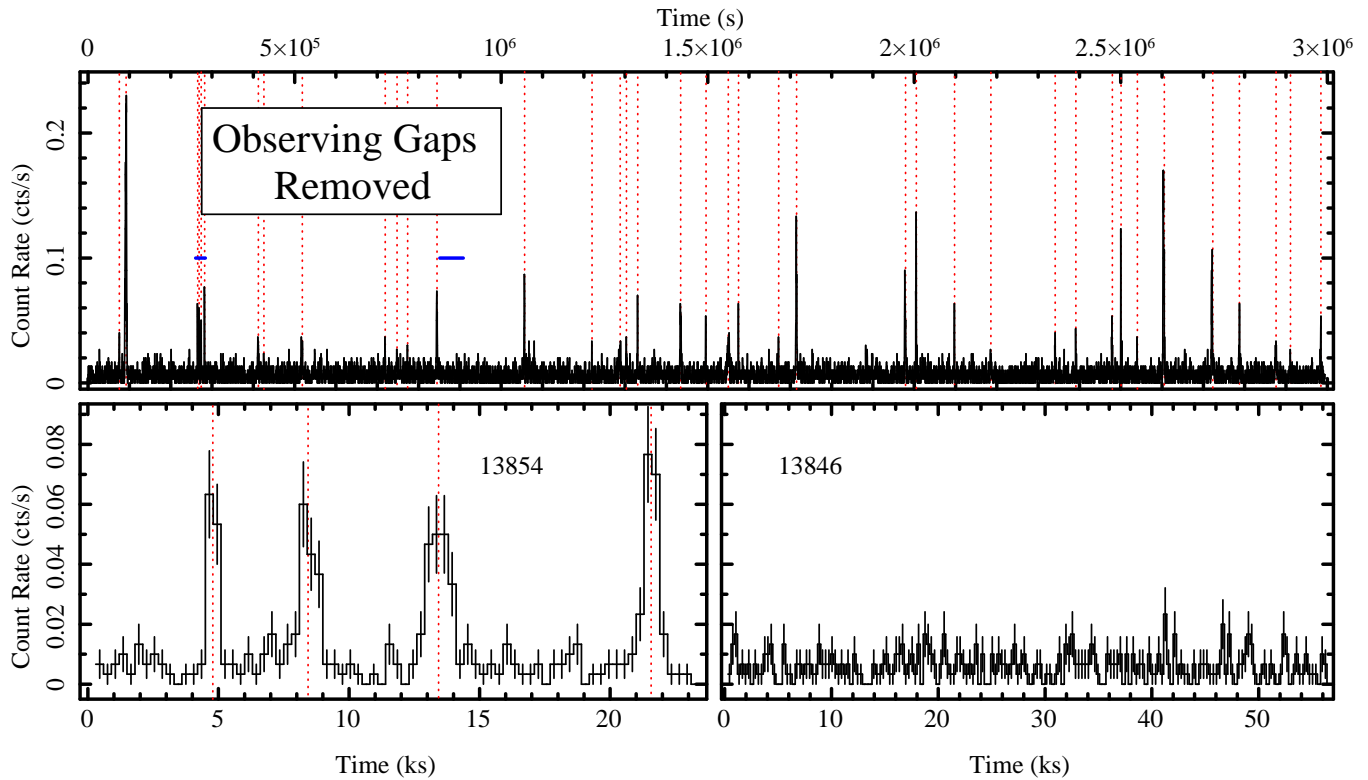}}
\caption{(Top): Combined zeroth and $1^{\rm st}$ order 2--8 keV \textit{Chandra} X-ray lightcurve of \sgra in 300 s bins for the entire 2012 XVP campaign \textit{with gaps removed}; events are extracted from a 2.5 pixel radius circular region around the source for the zeroth order and 5 pixel wide rectangular regions for the $1^{\rm st}$ order lightcurves. A number of flares of varying intensity are apparent, and are indicated by dotted red lines. Short blue horizontal lines indicate sample observations shown in the bottom panel. (Bottom): Sample lightcurves of an observation with (left) and without (right) detected flares. ObsID 13854 (left) shows four moderately bright flares within 20 ks.\label{fig:lc1}}
\end{figure*}

During the 2012 XVP campaign, \textit{Chandra} observed \sgra 38 times at high spectral resolution with the HETGS. The observations, detailed in Table \ref{tbl:obs}, took place between 2012 February 6 and October 29, with exposure times ranging from 14.53 ks to 189.25 ks. For thermal reasons, only five of the chips from the Advanced CCD Imaging Spectrometer-Spectroscopy array (ACIS-S) were used during the campaign (S0 was turned off). We reduce the data with standard tools from the {\sc ciao} analysis suite, version 4.5. The HETGS consists of two transmission gratings, the High Energy Grating (HEG) and the Medium Energy Grating (MEG), which disperse a fraction of the incoming photons across the detector. With five chips, each detected photon is recorded to a timing accuracy of 3.14 seconds, along with its energy as measured by ACIS and its wavelength and diffraction order as determined by the diffraction equation and the {\tt tg\_resolve\_events} tool. To study the X-ray variablility of \sgra, we extract 2--8 keV lightcurves of these events in 300-s bins, including only photons in the zeroth order (i.e.\ the undispersed photons) and the HEG and MEG $\pm1^{\rm st}$ orders (hereafter $1^{\rm st}$ order photons). All our subsequent timing analysis is done in the Interactive Spectral Interpretation System (ISIS; \citealt{HD00,Houck02}).

%\setcounter{figure}{0}
%\begin{figure*}
%\centerline{\includegraphics[width=\textwidth]{f1b}}
%\caption{Continued.\label{fig:lc2}}
%\end{figure*}
%\clearpage
 
As reported in \citet{Nowak12}, a number of X-ray flares are detected in these observations. It is an impossible task to distinguish the \sgra photons from the underlying diffuse emission. However, the short flare time scales and the absence of associated optically thin radio flares (Section \ref{sec:stat}; \citealt{Baganoff01}) argue that the flare emission arises from close to the black hole. For our purposes, then, it is sufficient to minimize the background by using small extraction regions as in \citet{Nowak12}: a $\sim2.5$ pixel (1.25\arcsec) radius circle for the zeroth-order events, 5 pixel wide rectangular regions for the $1^{\rm st}$ order photons, and grating order tolerances of $\pm0.2.$ Since much of the background emission is found in the $1^{\rm st}$ order (Section \ref{sec:stat}), a search of the zeroth order alone would also have good sensitivity to flares. But the $1^{\rm st}$ order is essential for pileup calibration (Section \ref{sec:pile}), and we prefer to use the same events to detect, characterize, and calibrate our flares.

\section{LIGHTCURVES AND FLARE DETECTION}
\label{sec:lc}
The resulting X-ray lightcurves are shown in Figure \ref{fig:lc1}. For observations spread over nine months and a number of spacecraft roll angles (which affects the diffuse emission in the grating extraction regions), there is remarkably little variation in the quiescent level (Table \ref{tbl:obs}). Highlighting the stability of the baseline emission are the numerous narrow flares, which appear in over half of the 2012 observations (Table \ref{tbl:obs}). Most apparent is the large flare early in the campaign, described in detail in \citet{Nowak12}, which is the brightest X-ray flare ever observed from \sgra (see also \citealt{Porquet03}). Figure \ref{fig:lc1} includes several comparably bright flares and numerous moderate and weak flares. One 2-day observation (ObsID 13840) contains no flares, which is consistent with the average rate of $\sim1$ flare per day (Section \ref{sec:discuss}).

In order to detect and characterize these X-ray flares, we use an algorithm based on direct fits to the X-ray lightcurves shown in Figure \ref{fig:lc1}. There are systematic uncertainties associated with any particular choice of search algorithm; the robustness of our search and an alternative method (the Bayesian Blocks routine; \citealt{Scargle12_arxiv}) are discussed in the Appendix. Here, we fit the 2--8 keV lightcurves with a model consisting of a constant baseline and Gaussian components to represent the flares. Because the count rates are small, we use the Cash statistic (\citealt{Cash79}) and the {\tt subplex} fit method.

After a first pass to estimate the baseline count rate, we perform an automated search for narrow flares on an observation-by-observation basis. In the algorithm, each time bin is examined: if the count rate is below the (fixed) quiescent level, it is ignored (see Section \ref{sec:flarevar} for a comparison of the data to a Poisson process). Otherwise, we add a ``flare'' at the center of the bin: a faint narrow Gaussian with an initial 1$\sigma$ width $\sigma_{\rm t}=150$ s. We fit for the flare amplitude and then allow $\sigma_{\rm t}$ to vary as well. We restrict $\sigma_t$ to be between 100 s and 1600 s (empirically-determined limits, below which the bin size starts to become large relative to the flare FWHM and above which confusion from nearby flares can interfere with the search process). If the resulting amplitude is larger than the quiescent level at 99\% confidence for a single trial, the Gaussian component is identified as a real flare, and we fit for the best amplitude, center, and width.

Because the brightest flare (\citealt{Nowak12}) showed marked asymmetry, we also perform a search for time substructure in the detected flares. Leaving the initial (significant) component free, we add additional ``subflares'' as above, with flare center times constrained to occur within $\pm2\sigma_{\rm t}$ of the main component, until additional substructure is no longer significant at the 90\% level. The substructure is equally likely to appear before and after the peak. Finally, once all flares and substructure have been identified, we calculate 90\% confidence limits for each parameter, including the background level. We define the start (stop) time of a flare to be the minimum (maximum) value of the $2\sigma_{\rm t}$ lower (upper) limits for all its subflares. For each flare, we tabulate the start and stop times, durations, background count rates, rise and decay times, and note whether the flare was truncated by the beginning or end of an observation. We use customized models to fit for the peak count rate and the fluence within the start and stop times of each flare directly (fitting for these parameters and their uncertainties is faster and more reliable than combining all the subflares and propagating their uncertainties).

\begin{deluxetable*}{cccccccccccc}
\tabletypesize{\scriptsize}
\tablecaption{2012 \textit{Chandra} XVP Observations of Flares from \sgra\label{tbl:obs}}
\tablewidth{0pt}
\tablehead{
\colhead{}  &
\colhead{Obs Start}  & 
\colhead{Exp.}  &
\colhead{Bkg.}  & 
\colhead{Flare} &
\colhead{Flare} &
\colhead{Fluence} &
\colhead{Peak Rate}  & 
\colhead{Duration}  & 
\colhead{\# Sub-}  &
\colhead{} &
\colhead{}  \\
\colhead{ObsID}  &
\colhead{(MJD)}  &
\colhead{(ks)}  &
\colhead{Rate}  & 
\colhead{Start} &
\colhead{Stop} &
\colhead{(cts)} &
\colhead{(cts/s)}  &
\colhead{(s)}  &
\colhead{Flares} & 
\colhead{$F_{2-8}^{\rm abs}$} & %($10^{-12}$ erg cm$^{-2}$ s$^{-1}$)} &
\colhead{$L_{2-10}^{\rm unabs}$}%($10^{34}$ erg s$^{-1}$)} 
}
\startdata
13850 & 55963.026 & 60.06 & $5.9\pm0.5$ & \nodata & \nodata & \nodata & \nodata & \nodata & \nodata &  \nodata & \nodata \\
14392 & 55966.262 & 59.25 & $6.2\pm0.6$ & 55966.433 & 55966.464 & $33^{+12}_{-11}$ & $0.021^{+0.011}_{-0.008}$ & 2600 & 1 & 0.8 & 1.7 \\
      &           &       &              & 55966.603 & 55966.666 & $706^{+46}_{-44}$ & $0.23\pm0.03$ & 5450 & 4 & 8.5 & 19.2 \\
14394 & 55967.136 & 18.06 & $6.2^{+1.0}_{-0.9}$ & \nodata & \nodata & \nodata & \nodata & \nodata & \nodata & \nodata & \nodata \\
14393 & 55968.426 & 41.55 & $7.7\pm0.7$ & \nodata & \nodata & \nodata & \nodata & \nodata & \nodata & \nodata & \nodata \\
13856 & 56001.365 & 40.06 & $5.6\pm0.6$ & \nodata & \nodata & \nodata & \nodata & \nodata & \nodata & \nodata & \nodata \\
13857 & 56003.373 & 39.56 & $6.8\pm0.7$ & \nodata & \nodata & \nodata & \nodata & \nodata & \nodata & \nodata & \nodata \\
13854 & 56006.425 & 23.06 & $5.7\pm1.0$ & 56006.486 & 56006.493 & $32^{+11}_{-9}$ & $0.06^{+0.03}_{-0.02}$ & 600 & 1 & 3.3 & 7.4 \\
      &           &       &              & 56006.524 & 56006.540 & $40^{+12}_{-10}$ & $0.05^{+0.02}_{-0.01}$ & 1350 & 1 & 1.8 & 4.1 \\
      &           &       &              & 56006.580 & 56006.599 & $49^{+13}_{-12}$ & $0.05^{+0.02}_{-0.01}$ & 1600 & 1 & 1.9 & 4.2 \\
      &           &       &              & 56006.678 & 56006.690 & $49^{+13}_{-14}$ & $0.07^{+0.03}_{-0.02}$ & 950 & 1 & 3.1 & 7.1 \\
14413 & 56007.281 & 14.72 & $6.1^{+1.1}_{-1.0}$ & \nodata & \nodata & \nodata & \nodata & \nodata & \nodata & \nodata & \nodata \\
13855 & 56008.476 & 20.06 & $6.7^{+1.0}_{-0.9}$ & \nodata & \nodata & \nodata & \nodata & \nodata & \nodata & \nodata & \nodata \\
14414 & 56009.742 & 20.06 & $6.3\pm0.9$ & \nodata & \nodata & \nodata & \nodata & \nodata & \nodata & \nodata & \nodata \\
13847 & 56047.678 & 154.07 & $6.0\pm0.3$ & 56048.510 & 56048.548 & $59^{+16}_{-14}$ & $0.03^{+0.02}_{-0.01}$ & 3250 & 3 & 1.1 & 2.5 \\
      &           &       &              & 56048.679 & 56048.693 & $15^{+8}_{-7}$ & $0.020^{+0.015}_{-0.010}$ & 1200 & 1 & 0.7 & 1.7 \\
14427 & 56053.834 & 80.06 & $6.8\pm0.5$ & 56054.107 & 56054.154 & $49^{+15}_{-13}$ & $0.03^{+0.02}_{-0.01}$ & 4050 & 2 & 0.7 & 1.6 \\
13848 & 56056.502 & 98.16 & $6.2\pm0.4$ & \nodata & \nodata & \nodata & \nodata & \nodata & \nodata & \nodata & \nodata \\
13849 & 56058.138 & 178.75 & $6.7\pm0.3$ & 56058.687 & 56058.705 & $24^{+10}_{-11}$ & $0.03\pm0.01$ & 1600 & 1 & 0.9 & 2.0 \\
      &           &       &              & 56059.006 & 56059.044 & $33^{+15}_{-11}$ & $0.017^{+0.010}_{-0.007}$ & 3250 & 1 & 0.6 & 1.4 \\
      &           &       &              & 56059.314 & 56059.329 & $21^{+10}_{-8}$ & $0.03^{+0.02}_{-0.01}$ & 1250 &1 & 1.0 & 2.3 \\
      &           &       &              & 56060.127 & 56060.168 & $124^{+21}_{-19}$ & $0.06\pm0.01$ & 3500 & 1 & 2.2 & 4.9 \\
13846 & 56063.445 & 56.21 & $6.0^{+0.6}_{-0.5}$ & \nodata & \nodata & \nodata & \nodata & \nodata & \nodata & \nodata & \nodata \\
14438 & 56065.187 & 25.79 & $6.2\pm0.8$ & \nodata & \nodata & \nodata & \nodata & \nodata & \nodata & \nodata & \nodata \\
13845 & 56066.446 & 135.31 & $6.3\pm0.4$ & 56067.863 & 56067.888 & $102^{+18}_{-17}$ & $0.08\pm0.02$ & 2150 & 1 & 2.9 & 6.6 \\
14460 & 56117.940 & 24.06 & $7.2\pm0.9$ & \nodata & \nodata & \nodata & \nodata & \nodata & \nodata & \nodata & \nodata \\
13844 & 56118.966 & 20.06 & $6.0^{+1.0}_{-0.8}$ & \nodata & \nodata & \nodata & \nodata & \nodata & \nodata & \nodata & \nodata \\
14461 & 56120.242 & 50.96 & $7.0\pm0.6$ & \nodata & \nodata & \nodata & \nodata & \nodata & \nodata & \nodata & \nodata \\
13853 & 56122.026 & 73.66 & $5.2\pm0.4$ & 56122.650 & 56122.656 & $ 8^{+6}_{-4}$ & $0.03^{+0.02}_{-0.01}$ & 500 & 1 & 1.0 & 2.3 \\
13841 & 56125.880 & 45.07 & $6.3^{+0.7}_{-0.6}$ & \nodata & \nodata & \nodata & \nodata & \nodata & \nodata & \nodata & \nodata \\
14465 & 56126.975 & 44.34 & $5.8\pm0.7$ & 56126.979 & 56127.038 & $58^{+16}_{-15}$ & $0.03^{+0.02}_{-0.01}$ & 5100\tablenotemark{a} & 2 & 0.7 & 1.6 \\
      &           &       &              & 56127.172 & 56127.202 & $26^{+11}_{-9}$ & $0.017^{+0.010}_{-0.007}$ & 2550 & 1 & 0.6 & 1.4 \\
14466 & 56128.526 & 45.08 & $7.0\pm0.7$ & 56128.549 & 56128.553 & $29^{+10}_{-9}$ & $0.06^{+0.03}_{-0.02}$ & 400 & 1 & 4.5 & 10.2 \\
13842 & 56129.495 & 191.74 & $6.1\pm0.3$ & 56130.182 & 56130.225 & $101^{+19}_{-17}$ & $0.06^{+0.03}_{-0.02}$ & 3700 & 4 & 1.7 & 3.8 \\
      &           &       &              & 56130.906 & 56130.921 & $46\pm13$ & $0.06\pm0.02$ & 1300 & 1 & 2.1 & 4.8 \\
      &           &       &              & 56131.494 & 56131.585 & $119^{+23}_{-21}$ & $0.027^{+0.010}_{-0.008}$ & 7800 & 2 & 0.9 & 2.1 \\
13839 & 56132.294 & 176.24 & $6.5\pm0.3$ & 56132.385 & 56132.399 & $38\pm12$ & $0.05\pm0.02$ & 1150\tablenotemark & 1 & 2.0 & 4.5 \\
      &           &       &              & 56133.512 & 56133.521 & $14^{+8}_{-7}$ & $0.03^{+0.02}_{-0.01}$ & 750 & 1 & 1.2 & 2.6 \\
      &           &       &              & 56133.997 & 56134.042 & $251^{+28}_{-26}$ & $0.14\pm0.03$ & 3950 &  2 & 3.9 & 8.9 \\
13840 & 56134.835 & 162.50 & $7.0^{+0.4}_{-0.3}$ & \nodata & \nodata & \nodata & \nodata & \nodata & \nodata & \nodata & \nodata \\
14432 & 56138.539 & 74.26 & $6.2\pm0.5$ & 56139.368 & 56139.417 & $166^{+24}_{-22}$ & $0.08^{+0.03}_{-0.02}$ & 4250\tablenotemark{a} & 2 & 2.4 & 5.4 \\
13838 & 56140.729 & 99.55 & $6.5\pm0.4$ & 56141.009 & 56141.035 & $135^{+21}_{-23}$ & $0.10\pm0.02$ & 2250 & 1 & 3.7 & 8.4 \\
13852 & 56143.109 & 156.55 & $6.2\pm0.3$ & 56143.314 & 56143.332 & $58\pm15$ & $0.06^{+0.02}_{-0.01}$ & 1550 & 1 & 2.3 & 5.1 \\
      &           &       &              & 56144.321 & 56144.363 & $33^{+14}_{-12}$ & $0.015^{+0.008}_{-0.006}$ & 3600 & 1  & 0.6 & 1.2 \\
14439 & 56145.928 & 111.72 & $5.9\pm0.4$ & 56147.131 & 56147.151 & $27^{+11}_{-10}$ & $0.03\pm0.01$ & 1750 & 1 & 0.9 & 2.1 \\
14462 & 56206.689 & 134.06 & $5.8\pm0.3$ & 56207.174 & 56207.194 & $30^{+11}_{-10}$ & $0.03^{+0.02}_{-0.01}$ & 1700 & 1 & 1.1 & 2.4 \\
      &           &       &              & 56208.187 & 56208.222 & $54^{+15}_{-13}$ & $0.05^{+0.03}_{-0.02}$ & 2950 & 3 & 1.1 & 2.5 \\
14463 & 56216.036 & 30.77 & $6.7\pm0.8$ & 56216.239 & 56216.248 & $58^{+14}_{-12}$ & $0.11^{+0.04}_{-0.03}$ & 750 & 1 & 4.7 & 10.7 \\
13851 & 56216.784 & 107.05 & $5.8\pm0.4$ & 56217.094 & 56217.098 & $15^{+8}_{-6}$ & $0.03^{+0.02}_{-0.01}$ & 400 & 1 & 2.4 & 5.4 \\
      &           &       &              & 56217.816 & 56217.884 & $372^{+34}_{-32}$ & $0.16^{+0.04}_{-0.03}$ & 5900 & 4 & 3.9 & 8.9 \\
15568 & 56218.372 & 36.06 & $6.9^{+0.8}_{-0.7}$ & \nodata & \nodata & \nodata & \nodata & \nodata & \nodata & \nodata & \nodata \\
13843 & 56222.667 & 120.66 & $6.1\pm0.4$ & 56223.384 & 56223.464 & $193^{+26}_{-24}$ & $0.08\pm0.01$ & 6900 & 2 & 1.7 & 3.8 \\
15570 & 56225.146 & 68.70 & $5.7\pm0.5$ & 56225.230 & 56225.263 & $63^{+16}_{-14}$ & $0.05\pm0.02$ & 2800 & 2 & 1.4 & 3.1 \\
14468 & 56229.988 & 146.05 & $5.5\pm0.3$ & 56230.288 & 56230.362 & $74^{+19}_{-18}$ & $0.020^{+0.007}_{-0.004}$ & 6350 & 1 & 0.7 & 1.6 \\
      &           &       &              & 56230.724 & 56230.734 & $13^{+8}_{-7}$ & $0.02^{+0.02}_{-0.01}$ & 900 & 1 & 0.9 & 2.0 \\
      &           &       &              & 56231.566 & 56231.592 & $54^{+14}_{-13}$ & $0.05\pm0.02$ & 2250 & 3 & 1.5 & 3.3 
      \enddata
\tablecomments{%2012 \textit{Chandra} XVP observations of \sgra, with flares and their properties.
All dates are reported in MJD (UTC). The background rate is reported in units of $10^{-3}$ counts s$^{-1}.$ The flare fluence, peak rate, and duration are determined from fits to the X-ray lightcurve in 300 s bins, as described in Section \ref{sec:lc}, and are raw measurements not corrected for pileup. The number of subflares is the number of Gaussian components required for each flare. $F_{2-8}^{\rm abs}$ and $L_{2-10}^{\rm unabs}$ are preliminary estimates of the pileup-corrected mean absorbed $2-8$ keV flux and mean unabsorbed $2-10$ keV luminosities of each flare, in units of $10^{-12}$ erg s$^{-1}$ cm$^{-2}$ and $10^{34}$ erg s$^{-1},$ respectively. These fluxes and luminosities are estimated by scaling from the brightest flare and the results of \citet{Nowak12}.}
\tablenotetext{a}{This flare is truncated by the beginning or end of an observation.}
\end{deluxetable*}
Using this algorithm, we detect a total of 39 flares in 21 observations; the remaining 17 observations appear to be consistent with quiescent emission and/or undetectable flares. We report the measured properties of the 39 detected flares in Table \ref{tbl:obs}. The flares range in duration from 400 s (our shortest allowed time) to about 7800 s, in fluence from $8$ counts to $706$ counts, and in peak count rate from $0.015$ counts s$^{-1}$ to $0.23$ counts s$^{-1}$. The fluences are divided roughly 3:2 between the 0$^{\rm th}$ and $1^{\rm st}$ orders. As discussed in \citet{Nowak12}, there is some ambiguity associated with the peak count rates, since there may be substructure in the flares on time scales shorter than the present binning (see their Figure 2). Strictly speaking, the peak rates reported here are peak rates on time scales of 300 s, and should be regarded as lower limits on the ``instantaneous'' peak count rate during the flare. We affirm \citeauthor{Nowak12}'s suggestion that the least ambiguous properties of a flare are its absorbed 2--8 keV fluence and mean flux, as measured above the quiescent/background emission.

For comparison to \citet{Nowak12}, in addition to the fluence, Table \ref{tbl:obs} also includes estimates of the mean absorbed 2--8 keV flux and the mean unabsorbed 2--10 keV luminosity of each flare for comparison to \citet{Nowak12}. We assume that all flares have the same $\Gamma=2$ power law X-ray spectrum (see Section \ref{sec:hr}), so that the flux and luminosity of a flare are proportional to its mean count rate\footnote{Especially for flares with relatively low fluence, where the observed counts may not adequately sample the intrinsic spectrum, the uncertainty associated with these scalings is likely large.} (defined as the pileup-corrected fluence divided by the duration). We normalize to the brightest flare, which \citet{Nowak12} found had a mean absorbed $2-8$ keV flux of $8.5\pm0.9\times10^{-12}$ erg cm$^{-2}$ s$^{-1}$, and an unabsorbed $2-10$ keV luminosity of $L_{\rm X} = 19.2^{+7.2}_{-3.7}\times10^{34}$ erg s$^{-1}.$ Only the luminosity and flux are corrected for photon pileup (see Section \ref{sec:pile}); the rest of the quantities (i.e.\ count rates and fluences) in Table \ref{tbl:obs} represent raw data.   

\section{FLARE STATISTICS}
\label{sec:stat}
At first glance, the flares appear to make a relatively minor contribution to the X-ray emission from the Galactic Center: the total duration of the observed flares is only 104.4 ks (a duty cycle of 3.5\%), and the integrated fluence of all 39 flares, $\sim3,400$ counts, is small relative to the $\sim22,000$ total counts in our extraction region. Much of the $1^{\rm st}$ order flux is diffuse background emission, however, and the zeroth order counts suggest that flares may have contributed as much as $\sim30\%$ of the total X-ray emission from the inner $1.25\arcsec$ during the 2012 XVP. This is precisely the ratio implied by a comparison of the summed energy of the observed flares, $E_f=5\times10^{39}$ erg, and the total energy emitted in quiescence $E_q\sim10^{40}$ erg (calculated from the steady quiescent luminosity, $L_q=3.6\times10^{33}$ erg s$^{-1}$; \citealt{Nowak12}). Again, we note that much of the steady quiescent X-ray emission could originate far from \sgra, near the Bondi radius. 

With the origin of both the flares and the quiescent emission (e.g.\ \citealt{Sazonov12}, but see \citealt{Wang13}) still unclear, in this section we focus on three questions of immediate interest: (1) are there multiple populations of flares, and if so, (2) how does the spectral hardness of the flares vary with their luminosity and duration, and (3) how much do undetected flares contribute to the baseline/quiescent emission of \sgra?

\subsection{Observational Biases}
Before we can address these questions, we must quantify their intrinsic properties. In the present study, this requires compensating our observed statistics for pileup, which tends to reduce observed count rates, as well as incompleteness and false detections in our flare-finding algorithm. 

\subsubsection{Pileup Correction}
\label{sec:pile}
Pileup occurs when two or more photons land in a single event detection cell (for \textit{Chandra}, a $3\times3$ pixel ``island'') during a single CCD frame; the resulting charge pattern on the CCD may be interpreted as a single energetic event, or it may be discarded. The net result is that pileup leads to reduced count rates and harder CCD spectra. The gratings, on the other hand, are pileup free up to relatively high count rates because the incident flux is dispersed over many more pixels. Thus we can use the $1^{\rm st}$ order lightcurves to correct for any possible pileup in the zeroth order count rates. 

To make this correction, we need to express the total observed count rate (i.e.\ the sum of the observed count rates in the zeroth and $1^{\rm st}$ orders, $r=r_0+r_1$) in terms of the incident rate $\Lambda_i$ and the $1^{\rm st}$ order rate $r_1.$ Note that in pileup calculations, all rates are expressed as counts per CCD frame (i.e.\ 3.1 s for five chips), not per unit time. Following \citet{Nowak12} Equation 2, $r$ can be written: 
\begin{equation}
\label{eq:pile}
r = \Lambda_i\left(1+\frac{K}{\alpha\Lambda_i}[\exp(\alpha\Lambda_i)-1]\exp(-\Lambda_i)\right)\left(\frac{r_1}{\Lambda_i}\right),
\end{equation}
where $K$ is a dimensionless function of the spectral shape; it is also the proportionality constant for Equation 2 in \citet{Nowak12}. $\alpha$ is the grade migration parameter in the pileup model: the odds of detecting $N$ piled photons as a single event is $\propto\alpha^{N-1}$ (\citealt{Davis01}). Here we assume $\alpha=1$ (\citealt{Nowak12}).   

The faintest flares in the 2012 XVP peak at observed count rates $<0.04$ counts s$^{-1}$ (see Section \ref{sec:demo}); the average count rate in these flares is $r_f\sim0.019$ counts s$^{-1}=0.062$ counts frame$^{-1}.$ In these faintest flares, we measure $r_1=0.52~r_f$; we suppose that pileup is negligible here, so that $r_{f}=\Lambda_{i}.$ Plugging these numbers into Equation \ref{eq:pile}, we find $K=0.94$ for the $\Gamma=2$ flares. With $K$ known, we can use Equation \ref{eq:pile} to calculate the pileup-corrected incident count rate $\Lambda_i$ for any observed count rate $r,$ as long as the incident radiation has approximately the same spectrum as the faint flares. With no significance evidence for variations in the flare spectrum with count rate (Section \ref{sec:hr}), this approximation is good enough for the purposes of pileup corrections. For each flare, then, we calculate pileup-corrected peak count rates and fluences (recalling that the fluence is the product of the mean rate and the duration\footnote{An alternative approach would involve integrating over the time-dependent pileup rate; the difference is well within our confidence limits.}). In all cases, the pileup correction is less than 20\%, consistent with the estimate for the brightest flare (\citealt{Nowak12}).
\begin{figure}
\centerline{\includegraphics[width=2.85in]{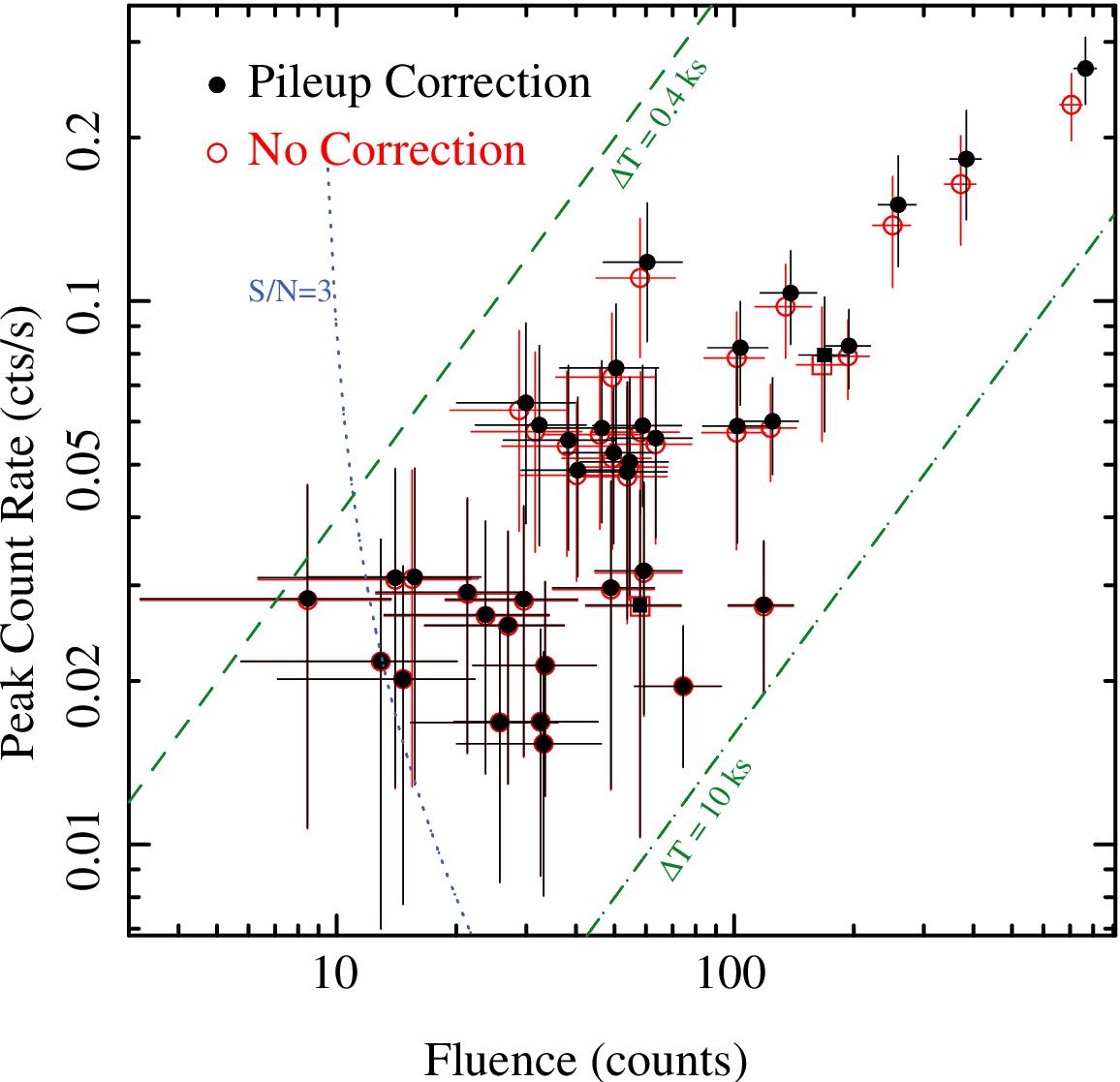}}
\centerline{\includegraphics[width=2.85in]{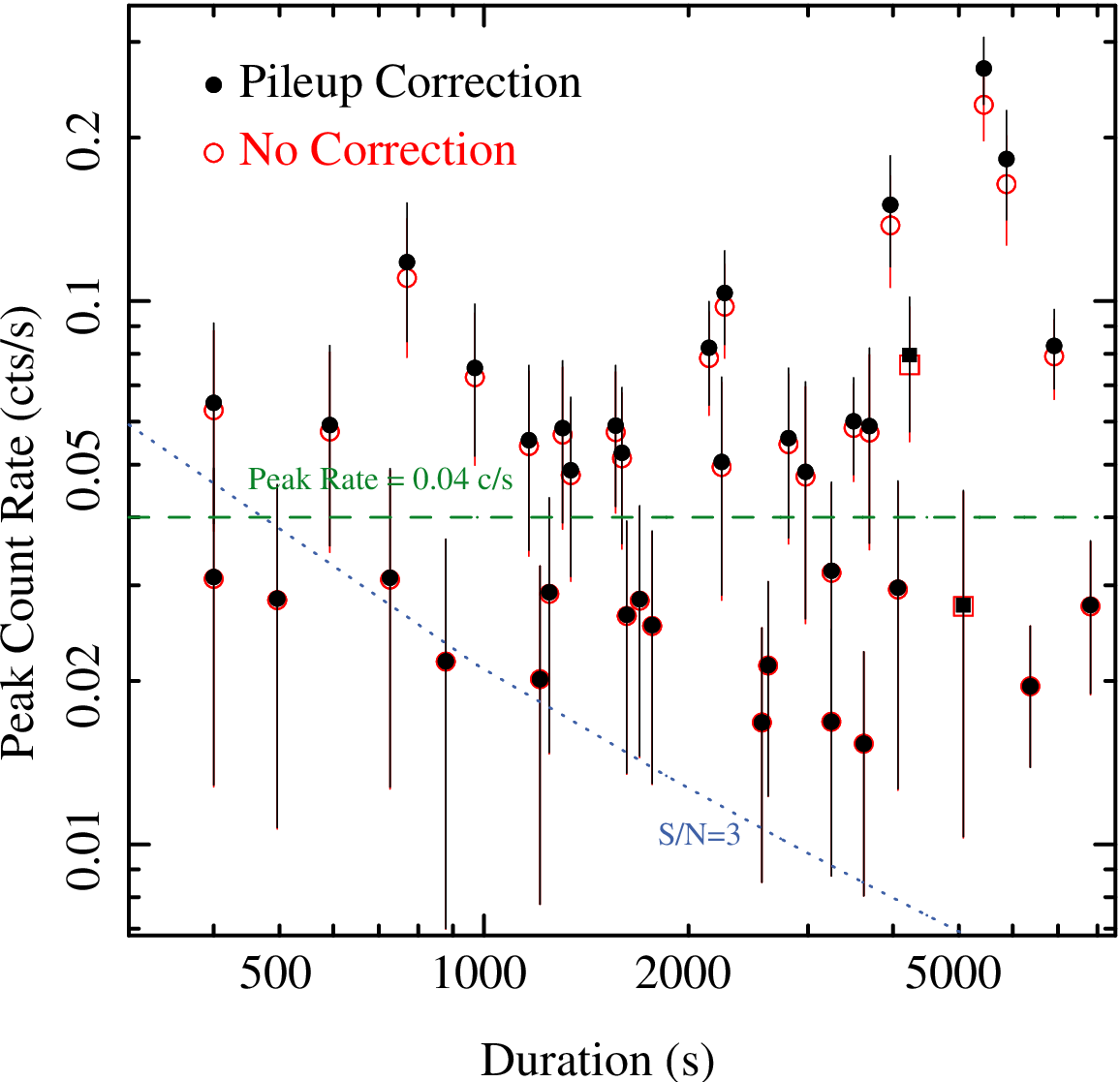}}
\centerline{\includegraphics[width=2.85in]{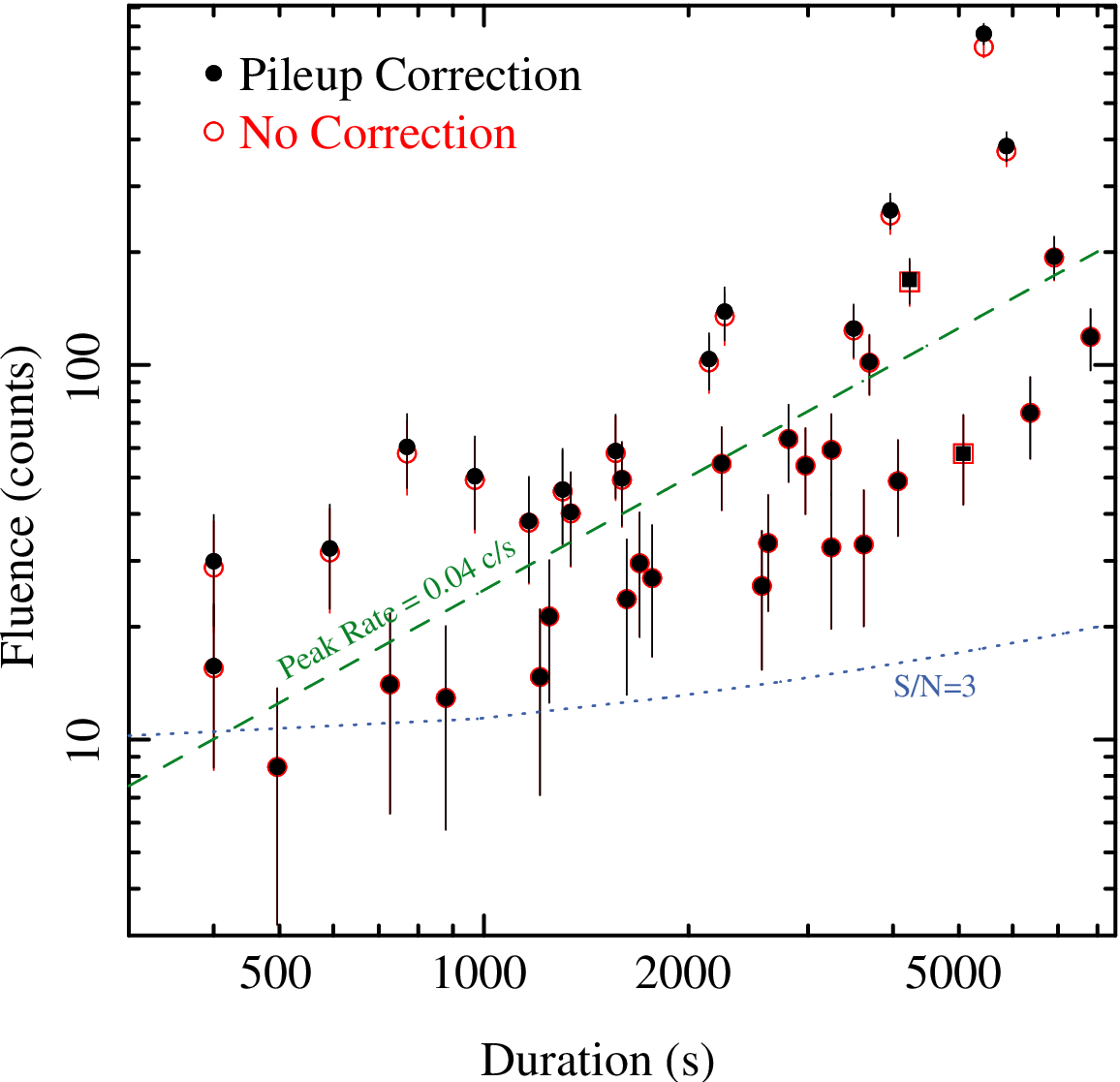}}
\caption{Relationships between the peak count rate, fluence, and duration of the 39 flares detected during the XVP. (Top): Peak count rate and fluence, with lines marking the expected relationship for 400 s flares, 10 ks flares, and flares with S/N$=3$. (Middle): Peak count rate and duration; the gap in peak rate around 0.04 counts s$^{-1}$ is apparent. (Bottom): Fluence and duration are strongly correlated. Squares denote flares that are truncated by the beginning/end of an observation.\label{fig:pr_fl_dt}}
\end{figure}

\subsubsection{Incompleteness and False Positives}
\label{sec:eff}
In order to assess the incompleteness of our sample and the false positive rate, we perform a suite of Monte Carlo simulations based on our fits to the observed lightcurves. To evaluate the false positive rate, we generate 100 Poisson-random realizations of the baseline emission in each observation and search for flares. No flares are detected, which suggests that our rate of identifying background fluctuations as flares is $<1\%$, and that all of our detected flares are likely real.

We quantify our flare detection efficiency $f$ with simulations in which we inject flares, generate Poisson-random realizations, search for flares, and use the detection rate to assess the incompleteness of our sample. For each observation, we simulate 500 lightcurves with randomly placed Gaussian flares: 25 flares for each of 20 different peak count rates (logarithmically-spaced between $0.01-0.3$ counts s$^{-1}$). To set the duration of each inserted flare, we randomly sample the observed duration distribution. When we search these simulated lightcurves, we define a flare inserted at time $t$ with width $\sigma_t$ as detected if our algorithm finds a flare in the time interval $t\pm3\sigma_t.$ With these results we can calculate $f$ for the binned distributions (see Section \ref{sec:dist}). At the low end, we have $f\gtrsim0.2-0.5;$ the incompleteness is not important for flares with fluences $\gtrsim80$ counts or peak rates above 0.04 counts s$^{-1}.$ To create efficiency-corrected histograms, we interpolate to find the detection efficiency at each observed flare, then take the sum of $f^{-1}$ in each bin (effectively dividing each histogram value by the average value of $f$ for the flares in that bin). We perform additional simulations to confirm (using our fluence distribution) that Poisson noise in the injected flares does not change our results to within the quoted errors.
 
\subsection{Flare Demographics}
\label{sec:demo}
In Figure \ref{fig:pr_fl_dt}, we present the relationships between the raw and pileup-corrected peak count rates, fluences, and durations of the 39 X-ray flares. We see moderate/strong correlations between the fluence of a flare and both its peak count rate (correlation coefficient $\rho\sim0.89$) and duration ($\rho\sim0.54$), but almost no correlation ($\rho\sim0.27$) between the peak count rate of a flare and its duration. Pileup does not have a significant effect on these correlations.

The top panel of Figure \ref{fig:pr_fl_dt}, shows the peak count rate versus fluence of the \sgra flares. Most of the flares are clustered around fluences of $\sim60$ counts, with peak count rates in the range $0.02-0.1$ counts s$^{-1}.$ At high fluence, the distribution of flares appears to narrow considerably. This does not appear to be an issue of incompleteness. For reference, we overplot a line representing a S/N ratio of 3, as well as lines of constant duration for single Gaussian flares with $1\sigma$ widths of 100 s (our minimum allowed width) and 2500 s, which correspond to flare durations of 400 s and 10 ks. We expect good sensitivity to flares inside this region. Thus the narrow distribution at high fluence is likely physical, and may indicate that the brightest flares have a preferred time scale ($\tau\sim4-6$ ks). The brightest flares seen by \textit{XMM-Newton} last $\sim3$ ks (\citealt{Porquet03,Porquet08}), so further study is merited.

\begin{deluxetable}{lcccc}
\tabletypesize{\scriptsize}
\tablecaption{Fits to Flare Distributions\label{tbl:dist}}
\tablewidth{0pt}
\tablehead{
\colhead{Parameter}  &
\colhead{Fluence}  & 
\colhead{Peak Rate}  &
\colhead{Duration}  &
\colhead{Luminosity}
}
\startdata
\multicolumn{4}{l}{\vspace{2mm}Powerlaw}\\
\vspace{0.5mm}$N$ & $67_{-34}^{+61}$  & $0.75_{-0.6}^{+1.6}$ & $5_{-4}^{+21}$ & $50\pm20$\\
\vspace{0.5mm}$\Gamma$ & $-1.5\pm0.2$ & $-1.9_{-0.5}^{+0.4}$ & $-0.9\pm0.2$ & $-1.9_{-0.4}^{+0.3}$  \\
\vspace{0.5mm}$\chi^{2}/\nu$ & 8.18/5 & 9.7/5 & 3.3/5  & 4.6/5\\
%\hline \\ [-1.5ex]
%\multicolumn{4}{l}{\vspace{2mm}Broken Power Law}\\
%\vspace{0.5mm}$N\tablenotemark{b}$ & $7_{-7}^{+47}$  & $0.6_{-0.5}^{+9.5}$ & $0.8_{-0.8}^{+698}$ & $50_{-20}^{+4050}$\\
%\vspace{0.5mm}$\Gamma_1$ & $-0.7_{-0.6}^{+0.7}$ & $-2.0_{-0.5}^{+0.4}$ & $-0.6_{-0.3}^{+0.4}$ & $<$-$1.2$ \\
%\vspace{0.5mm}Break\tablenotemark{a} & $50_{-20}^{+200}$  & $0.2_{-0.2}^{+0.09}$ & $4100_{-3600}^{+2600}$ & $13_{-12}^{+8}$\\
%\vspace{0.5mm}$\Gamma_2$ & $-2.0_{-0.9}^{+0.4}$ & $1_{-11}^{+9}$ & $<$-$0.8$ & $-5_{-5}^{+15}$ \\
%$\chi^{2}/\nu$ & 2.4/3 & 9.3/3 & 0.5/3 & 3.8/3  \\
\hline \\ [-1.5ex]
\multicolumn{4}{l}{\vspace{2mm}Cutoff Power Law}\\
\vspace{0.5mm}$N$ & $14_{-12}^{+44}$  & $0.7_{-0.6}^{+1.2}$ & $0.05_{-0.05}^{+7.6}$ & $46_{-15}^{+26}$\\
\vspace{0.5mm}$\Gamma$ & $-0.9_{-0.5}^{+0.8}$ & $-1.9_{-0.5}^{+0.8}$ & $-0.1_{-0.8}^{+1.0}$ & $-1.7_{-0.6}^{+1.1}$ \\
\vspace{0.5mm}Cutoff\tablenotemark{a,b} & $160_{-110}^{+780}$  & $>0.05$ & $>1500$ & $>4$\\
$\chi^{2}/\nu$ & 3.8/4 & 9.7/4 & 0.8/4 & 4.5/4  
\enddata
%\tablenotetext{b}{The broken power law normalization is such that below the break, $y=N x^{\Gamma_1}.$ }
\tablenotetext{a}{In units of counts, counts s$^{-1}$, s, and $10^{34}$ erg s$^{-1}$ for the fluence, peak rate, duration, and luminosity distributions, respectively.}
\tablenotetext{b}{The cutoff power law is $N x^{\Gamma}\exp(-x/$Cutoff).}
\tablecomments{Power law and cutoff power law fits to the distributions in Figure \ref{fig:hist}. $N$ is the power law normalization and $\Gamma$ is the power law index. Errors are 90\% confidence limits for a single parameter. Errors on $N$ are typically large because they involve extrapolating outside the domain of the data.}
\end{deluxetable}

\begin{figure}
%\centerline{\includegraphics[width=3.2in]{f6}}
%\centerline{\includegraphics[width=3.2in]{f7}}
%\centerline{\includegraphics[width=3.2in]{f5}}
\centerline{\includegraphics[width=3.15in]{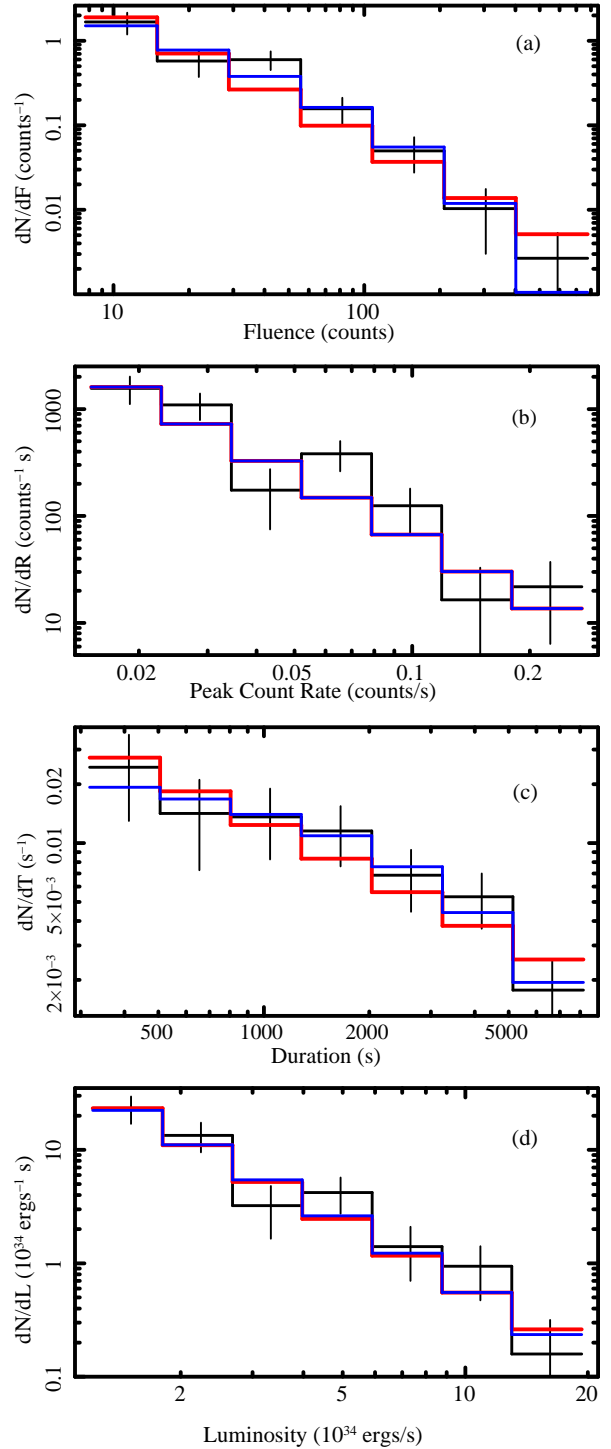}}
\caption{Distributions of flare fluence (a), peak count rate (b), duration (c), and mean luminosity (d). In all panels, the black histograms represent distributions corrected for both pileup and our detection efficiency, with Poisson errors on the number of flares in each bin. The red curves are the best power law fit, and the blue curves are the best cutoff power law fit.\label{fig:hist}}
\end{figure}

\subsection{Flare Distributions}
\label{sec:dist}
The distributions of the flare properties may also provide clues to their physical origin. In Figure \ref{fig:hist}, we present the binned differential distributions of the flare fluence, peak count rate, duration, and luminosity (all corrected for detection efficiency and for pileup; Sections \ref{sec:pile} and \ref{sec:eff}). These are our best assessment of the intrinsic flare properties in the observed range of parameters. Assuming Poisson uncertainties on each histogram bin, a power law model provides a good fit to the distributions of durations and luminosities (red curves in Figure \ref{fig:hist}; see Table \ref{tbl:dist} for fit parameters). The luminosity distribution falls off as $dN/dL\propto L^{-1.9},$ while the duration distribution falls off as $dN/dT\propto T^{-0.9}.$ The fluence and peak rate distributions, however, are not as well described by a single power law (the best fits, $dN/dF\sim F^{-1.5}$ and $dN/dR\sim R^{-1.9}$, lead to $\chi^{2}_{\nu}\sim1.6$ and $1.9$; see Table \ref{tbl:dist}). As an alternative, we try a high-end cutoff power law, shown in blue in Figure \ref{fig:hist}. This model provides an improved fit to the fluence distribution, but not the peak rate distribution (see Section \ref{sec:hr}). Based on the properties of the brightest flare (\citealt{Nowak12}), the best fit fluence cutoff of $\sim160$ counts corresponds to an absorbed $2-8$ keV fluence of $10^{-8}$ erg cm$^{-2}$ and a total $2-10$ keV energy of $2\times10^{38}$ erg. Although the additional parameter is clearly not required to describe the duration distribution, the best fit cutoff is about $3300$ s, which is similar to the orbital period in the inner accretion flow around \sgra. Neither of these cutoffs is well constrained.

An observational reason to prefer the models with a cutoff, however, is the absence of longer and brighter flares. For example, starting at our highest bins and integrating the best fit luminosity and duration power laws from $L=1.925\times10^{35}$ erg s$^{-1}$ to $L=10^{37}$ erg s$^{-1}$ and from $T=8175$ s to $T=10^{5}$ s, we would have expected to see $3.8_{-2.8}^{+6.0}$ very luminous flares and $52_{-28}^{+48}$ very long flares if the power laws continued indefinitely. The cutoffs reduce these numbers by factors of 2.3 and 14, respectively. Presently the result is marginal, but incorporating the two dozen previously observed \textit{XMM-Newton} and \textit{Chandra} flares will help determine the range over which the flare properties are distributed as power laws.

\subsubsection{Gaps and X-ray Colors}
\label{sec:hr}
One reason for the poor quality of the fits to the peak count rate distribution is an apparent gap near $R\sim0.04$ counts s$^{-1}$ (Figure \ref{fig:pr_fl_dt}). The gap is also apparent as a dip in $dN/dR$ in Figure \ref{fig:hist}. Using the same scaling as in Table \ref{tbl:obs}, the corresponding peak X-ray luminosity is roughly $5.4\times10^{34}$ erg s$^{-1}.$ There may also be a narrow gap around durations of $\sim2$ ks that appears in the characteristic time scale (fluence divided by peak rate) of the flares, as well as the flare durations as measured by the Bayesian Blocks method (see Appendix).

These gaps are particularly interesting given our search for multiple flare populations. There are many ways to test the statistical significance of a gap in a distribution (or, alternatively, whether the distribution is bimodal). One such test is the critical bandwidth test (\citealt{Silverman81,Minotte97,Hall01}). The critical bandwidth $h_{\rm crit}$ is defined as the smallest smoothing parameter for which the Gaussian kernel density estimate $f_{h_{\rm crit}}$ has a single mode. The test involves drawing sample data from this kernel density distribution, finding new critical bandwidths $\hat{h}_{\rm crit}$ for the new samples, and comparing them to $h_{\rm crit}.$ In general, if the data drawn from the smoothed distribution frequently require more smoothing than the original dataset, then the distribution is likely bimodal. Quantitatively, as described by \citet{Hall01}, the bimodality is significant at the level $\alpha$ if for an appropriately-chosen quantity $\lambda_\alpha,~P(\hat{h}_{\rm crit}/h_{\rm crit}>\lambda_\alpha)>1-\alpha.$ 

Given the large range of observed peak rates, we perform this test in log space and find that $h_{\rm crit}=0.163,$ which is $\sim50\%$ larger than the average $1\sigma$ logarithmic uncertainty on the data. Using the \citet{Hall01} formula for $\lambda_\alpha(\alpha),$ we estimate that the gap in the peak rate distribution is significant at the $\sim90\%$ level. The test does not specifically include measurement errors, but we find a similar significance level if we randomly sample the uncertainties instead of drawing new data from the kernel density distribution. Hereafter, we refer to flares above this gap as ``bright" flares; those below the gap are ``faint." Although the duration gap is not significant at the 90\% level in the 39 durations as reported in Table \ref{tbl:obs}, it appears in the Bayesian Blocks measurements and 2 ks is a reasonable time scale for dividing ``long" and ``short" flares..

\begin{comment}
\begin{deluxetable*}{cccccccc}
\tabletypesize{\scriptsize}
\tablecaption{Flare Types and Hardness Ratios}
\tablewidth{0pt}
\tablehead{
\colhead{}  &
\colhead{All}  & 
\colhead{Events}  &
\colhead{}  & 
\colhead{}  &
\colhead{1$^{\rm st}$-ord.} &
\colhead{} &
\colhead{1$^{\rm st}$/bkg.}  \\
\colhead{Type}  &
\colhead{HR}  & 
\colhead{Type}  &
\colhead{HR}  & 
\colhead{Type}  &
\colhead{HR} &
\colhead{Type} &
\colhead{HR}
}
\startdata
\vspace{1mm}Quiet & $0.77\pm0.01$ & Quiet & $0.77\pm0.01$     & Quiet & $0.55\pm0.01$ & Quiet  & \nodata\\
\vspace{1mm}Short & $1.7\pm0.1$      & Faint & $1.443\pm0.095$ & Faint & $0.94\pm0.08$ &  Faint & $1.4\pm0.3$\\ 
\vspace{2mm}Long & $1.82\pm0.07$ & Bright & $1.92\pm0.07$  & Bright & $1.2\pm0.05$ &  Bright & $1.4\pm0.1$\\
P$_{\rm K-S}$ & 0.61 &  & $2.4\times10^{-3}$ & & $0.13$ &  & \nodata\\
P$_{\rm HR}$ & 0.13 &  & $4\times10^{-5}$ & & $4.2\times10^{-3}$ &  & $0.52$
\enddata
\tablecomments{Hardnesses and stuff}
\end{deluxetable*}
\end{comment}

\begin{deluxetable}{lccc}
\tabletypesize{\scriptsize}
\tablecaption{Flare Types and Hardness Ratios\label{tbl:hr}}
\tablewidth{0pt}
\tablehead{
\colhead{Type}  &
\colhead{HR\tablenotemark{a}}  & 
\colhead{Type}  &
\colhead{HR\tablenotemark{a}}  
}
\startdata
\multicolumn{4}{c}{\vspace{2mm}Raw Events\tablenotemark{b}}\\
%\vspace{0mm}Quiet & $0.77\pm0.01$ & Quiet & $0.77\pm0.01$\\
\vspace{0mm}Short & $1.7\pm0.1$      & Faint & $1.443\pm0.095$\\
\vspace{1mm}Long & $1.82\pm0.07$ & Bright & $1.92\pm0.07$  \\
P$_{\rm K-S}$ & 0.61 &  & $2.4\times10^{-3}$ \\
P$_{\rm HR}$ & 0.13 &  & $4\times10^{-5}$\vspace{1mm}\\
\hline \\ [-1.5ex]
\multicolumn{4}{c}{\vspace{2mm}No Pileup ($1^{\rm st}$ order only)}\\
%\vspace{0mm}Quiet & $0.77\pm0.01$ & Quiet & $0.77\pm0.01$\\
\vspace{0mm}Short & $1.0\pm0.1$      & Faint & $0.94\pm0.08$\\
\vspace{1mm}Long & $1.1\pm0.06$ & Bright & $1.2\pm0.05$  \\
P$_{\rm K-S}$ & 0.45 &  & 0.13 \\
\vspace{1mm}P$_{\rm HR}$ & 0.18 &  & $4.2\times10^{-3}$ \\
\hline \\ [-1.5ex]
\multicolumn{4}{c}{\vspace{2mm}Bkg Subtracted, No Pileup}\\
%\vspace{0mm}Quiet & $0.77\pm0.01$ & Quiet & $0.77\pm0.01$\\
\vspace{0mm}Short & $1.27\pm0.25$      & Faint & $1.4\pm0.3$\\
\vspace{1mm}Long & $1.4\pm0.1$ & Bright & $1.4\pm0.1$  \\
%P$_{\rm K-S}$ & \nodata &  & \nodata \\
P$_{\rm HR}$ & 0.29 &  & 0.53 
\enddata
\tablecomments{This shows the effects of pileup and contamination by diffuse X-ray emission on HR. $P_{\rm K-S}$ and $P_{\rm HR}$ are the probabilities that the two types of flare differ, as described in Section \ref{sec:hr}. The top section reports HR based on all events extracted from the zeroth and $1^{\rm st}$ orders during the flares. The middle and bottom sections use only the $1^{\rm st}$ order events to avoid pileup; we subtract an estimate of the background emission in the bottom section (see text for details). After these corrections, no significant HR variations are detected.}
\tablenotetext{a}{Events extracted from from different extraction regions have different spectral responses, so their HR values should not be compared directly.}
\tablenotetext{b}{For reference, the raw background events have HR$=0.77\pm0.01.$ Thus the flares are significantly harder than the quiescent emission.}
\end{deluxetable}

In \citet{Nowak12}, we found that the brightest flare (ObsID 14392) appeared to be harder than its local background emission, and a Kolmogorov-Smirnov (K-S) test on the extracted events implied that the color difference between the flare and non-flare intervals was significant at a level of at least 95\% (with a maximum $p$-value of $P=4.6\times10^{-2}$ if we considered only zeroth order photons). With 38 more flares, $\sim2500$ additional flare photons, and much-improved statistics, we confirm that the flares are harder (see below and Table \ref{tbl:hr}); the probability that the flares and the quiescent emission have the same spectrum is indistinguishable from zero, even if we consider only the zeroth or $1^{\rm st}$ order photons alone. We are therefore confident that the X-ray spectrum of \sgra is harder during flares than in quiescence. Detailed spectral fits to all the observed flares will follow in future work, but here we focus on the X-ray colors of the flares above and below the peak rate and duration gaps. 

We measure the X-ray color with a hardness ratio (HR), which we define as the ratio of the $4-8$ keV counts to the $2-4$ keV counts (we calculate uncertainties on HR assuming square-root uncertainties on the numerator and denominator). The comparison of hardness ratios for different types of flares is detailed in Table \ref{tbl:hr}. We calculate the significance of the apparent differences in HR in two ways: (1) a K-S test on the relevant event lists, and (2) a Monte Carlo estimate of the probability that the two HR values are equal given their errors (which we assume to be Gaussian). For example, a K-S test indicates that the observed energy spectrum during faint flares differs from the observed spectrum during bright flares at the $\sim99.7\%$ level. Our HR-based estimate (using 1 million trials) implies an even higher level of significance, which could indicate that the errors on HR are underestimated or not Gaussian, or it could be due to the fact that it is easier to tell the difference between two numbers than two sets of many numbers: by using HR, we have condensed all the available spectral information into a single quantity.

There are two important caveats to these estimates, both of which we have alluded to above: pileup and the background/diffuse X-ray emission. Despite the null result (below), we believe it is instructive to explore the quantitative influence of these effects, especially because of their importance for future detailed spectral analysis of \sgra. Pileup affects the CCD spectrum at high fluxes, and may therefore bias the bright flares towards increased hardness. To account for this in Table \ref{tbl:hr}, we perform the same calculations as above using only the 1${\rm st}$ order events (which are free of pileup; Section \ref{sec:stat}). Based on the results, it does not appear that pileup is a major source of bias. Part of the reason for this may be that the flare spectrum (\citealt{Nowak12}) peaks around 4 keV, where our energy bands are delineated. Piled photons from the $2-4$ keV band will therefore be detected in the $4-8$ keV band, while piled photons in the harder band are not detected at all. So, to lowest order, pileup may therefore not result in a large change in HR for the bright flares.

One of the merits of the HR method over the K-S test is that it allows a rough subtraction of the underlying X-ray background, which turns out to be an important factor, since it is much softer than the flares and makes a larger fractional contribution to the faint flares. We use the rate of background events above and below 4 keV to estimate the expected number of background/diffuse photons during each type of flare. We then subtract these events from the appropriate energy band and recalculate HR; the results are shown in the bottom section of Table \ref{tbl:hr}. Considering only the unpiled, background-subtracted $1^{\rm st}$ order photons, there is no evidence for any HR changes between bright and faint flares or long and short flares. 

This leaves the significance of the gaps up for debate. If flares above and below these gaps had different spectral properties, we might be more inclined to view the gaps as real gaps separating two distinct emission mechanisms. At present, we have no evidence for that conclusion, nor can we rule out  a statistical fluctuation or a single flare mechanism with multiple characteristic time scales or luminosities.

%brightest fluence absorbed 4.7e-8 erg / cm^2, 706 cts -> 6.65e-11 erg/cm^2/ct
%brightest energy unabsorbed 1e39, 706 cts -> 1.4e36 erg/ct

% 155.37 counts corresponds to 1e-8 erg/cm^2, 2e38 erg = 2e14 kg

% flux in quiescence is 0.147e-12 erg/s/cm^2.
% luminosity in quiescence is 0.36e34 erg/s/cm^2.
% total emitted energy in quiescence is 1e40 erg

% the total flare luminosity times the flare duration is 4.7e39 erg

\subsection{Flares and the Quiescent Emission}
Because the origin of the quiescent X-ray emission from \sgra is still not well understood, and because the X-ray flares contribute one-third of the observed radiation from the inner $1.25\arcsec$ in the 2012 XVP, it is worth asking whether weak flares could make a significant contribution to the \textit{quiescent} emission from \sgra. Specifically, given the observed fluence distribution, what fraction of the quiescent emission could be attributed to undetected flares?

Our ability to answer this question robustly is limited by the (required) assumption that the fluence distribution can be extrapolated significantly below the observed fluence range. However, because the power law and cutoff power law fits differ significantly in slope at the low fluence end, $\Gamma=-(0.9-1.5)$, a comparison of the undetected fluence in these models should be indicative of the overall uncertainty on the extrapolation to low fluence. 

To estimate the integrated fluence of undetected flares, we modify our model normalizations to be the integral of $F(dN/dF)$ from $10^{-4}$ counts to $\sim8.5$ counts (the weakest detected flare). In the power law model, we find a total of $390_{-180}^{+380}$ counts below our detection limit; applying the same energy scalings as above, this corresponds to a fluence of $\sim2.6\times10^{-8}$ erg cm$^{-2}$ and a total energy of $5.5\times10^{38}$ erg, or $5.1\%$ of the steady X-ray output from the inner $1.25\arcsec$ during the XVP (the 90\% upper limit is 10\%). Note that here we are comparing background-subtracted fluences and background-subtracted X-ray emission, so the implication is that very weak flares do not contribute much more than 10\% of the quiescent luminosity. 

Indeed, the constraint is much tighter if we consider the cutoff power law model, in which there are $130_{-90}^{+220}$ counts below the detection limit (a fluence of $10^{-8}$ erg cm$^{-2}$ and a total energy of $2.2\times10^{38}$ erg, with a 90\% upper limit to the fractional flare contribution of $2.9\%).$ This is itself a slight overestimate, since for simplicity we calculate the integral as though there is no cutoff (which, at 160 counts, is well above the detection limit of 8.5 counts). If the distribution of very weak flares differs substantially from that of our detected flares, then it would be possible for the contribution to be greater. %However, we believe 5--10\% is a reasonably robust upper limit, given that the best fit cutoff power law model implies a contribution that is a factor of several smaller (see Section \ref{sec:discuss} for a discussion of the significance of this number). 

\subsubsection{Flares and Quiescent Variability}
\label{sec:flarevar}
If weak flares contribute as much as 10\% of the underlying X-ray emission, this should be apparent in the variability of the quiescent lightcurve. Is this excess variability present, and is it consistent with white noise? A stringent upper limit on correlated variability might imply that the X-rays come from an extended accretion flow or many independent sources (e.g.\ a star cluster; \citealt{Sazonov12}; see also \citealt{Wang13}).

\begin{figure}
\centerline{\includegraphics[width=3.2in]{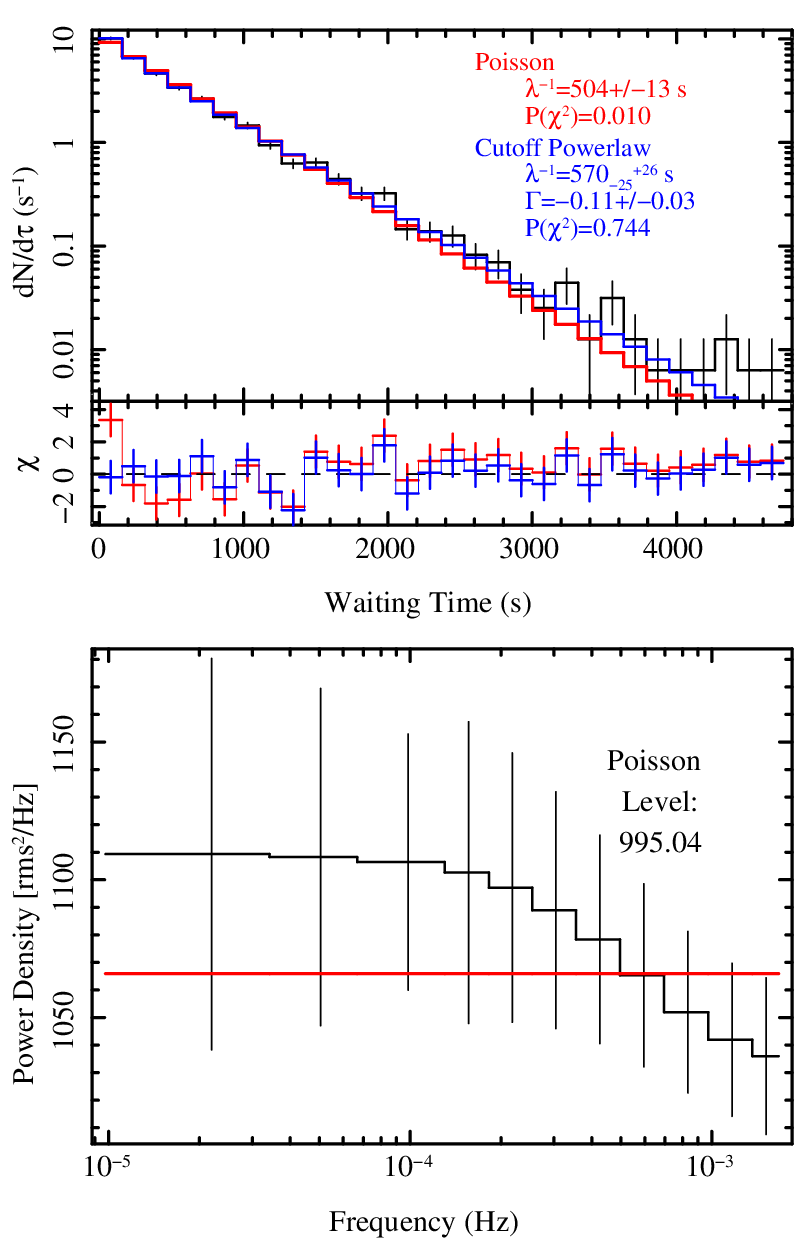}}
\caption{Waiting times (top) and simulated power spectrum (bottom) of the quiescent emission from \sgra. We fit the waiting time distribution with an exponential model (red) and a power-law-times-exponential model (blue). The simulated power spectrum is statistically indistinguishable from white noise, and is shown with the best fit constant model. See text for details. \label{fig:poisson}}
\end{figure}

If weak flares do not contribute to the X-ray variability, the waiting times between photons should be described by an exponential distribution. In order to avoid contamination by the diffuse background emission (see Section \ref{sec:hr}), we consider only quiescent photons detected in the zeroth order. We find 5,480 pairs of such photons that are not separated by either flares or gaps between observations. The resulting distribution of waiting times is shown in the top panel of Figure \ref{fig:poisson}; we assume $\sqrt N$ errors on each bin. A fit with an exponential curve is not formally acceptable ($\chi^{2}_{\nu}=1.72$ for 28 degrees of freedom), but suggests a characteristic rate $\lambda=0.002$ s$^{-1},$ or a characteristic time scale of $\lambda^{-1}=504\pm13$ s (90\% errors). A cutoff power law provides a better fit, with $\lambda^{-1}=570_{-25}^{+26}$ s and a power law slope of $-0.11\pm0.03.$ In effect, this means that there are fewer short wait times than expected for a Poisson process (or more long wait times). This suggests some correlated variability on short time scales, perhaps due to undetected flares.

Nevertheless, it is difficult to visualize correlated waiting times, so we have also used the cutoff power law model to create a simulated power spectrum (since the observed lightcurve is contaminated by flares). Assuming the counts in each histogram bin are Poisson-distributed, we generate 10,000 new waiting time distributions and fit each with the same model. We use the resulting fit models to generate new sets of waiting times, from which we build simulated event arrival times and lightcurves. From these, we calculate binned power spectra\footnote{The normalization of each power spectrum is such that its integral over frequency gives the squared fractional RMS variability; pure Poisson noise would have a power spectrum independent of frequency at a level $2/r,$ where $r$ is the mean count rate (see \citealt{Nowak99a} and references therein).}, taking the errors on each frequency bin to be the sample standard deviation of all 10,000 power spectra in that bin. We ignore all frequencies below $\sim10^{-5}$ Hz (comparable to the mean value of one over the observation length), so as not to give overmuch weight to frequencies that are not well represented in our data. Still, the errors are quite large, so we rebin the four lowest-frequency bins by a factor of four, and the next four bins by a factor of two.  The resulting power spectrum (bottom panel of Figure \ref{fig:poisson}) should be a decent approximation of the intrinsic variability of the inner $1.25\arcsec,$ free of contamination by strong flares and observational gaps. 

We can draw several conclusions from this analysis. First, Poisson noise dominates the power in the plotted frequency range. But there is also a $\sim10\%$ variability component above pure Poisson noise (see also Sections \ref{sec:dist} and \ref{sec:discuss}). Within observational uncertainties, this component is consistent with uncorrelated noise, but is also well described by a power law ($P_{\nu}\propto\nu^{-0.2}$) or a zero-frequency Lorentzian with a FWHM of $\sim10^{-3}$ Hz. The time scale is comparable to the orbital period at the innermost stable circular orbit. The shape of the power spectrum is similar to what is seen in many X-ray binaries and AGN (e.g.\ \citealt{Nowak99a,Uttley02,McClintockRemillard06,Wilkinson09}), but could also be produced by the superposition of many weak exponential flares with a characteristic duration of $\sim1$ ks (e.g.\ \citealt{Aschwanden11} and references therein). This result, a $\sim10\%$ excess of power that is consistent with short, weak flares occurring on a range of time scales, corroborates our findings based on extrapolating the flare distribution to low fluence.

%\begin{figure}
%\centerline{\includegraphics[width=3.2in]{f6gs}}
%\centerline{\includegraphics[width=3.2in]{f7gs}}
%\centerline{\includegraphics[width=3.2in]{f5gs}}
%\caption{\label{fig:duration}}
%\end{figure}

\section{DISCUSSION}
\label{sec:discuss}

In three megaseconds of observations during 2012, \textit{Chandra} detected a total of 39 X-ray flares from \sgra, spanning a factor of $\sim20$ in average luminosity (Table \ref{tbl:obs}), with the brightest flare topping out at a peak luminosity of $\sim5\times10^{35}$ erg s$^{-1}$ (\citealt{Nowak12}). Most of the observed flares have moderate luminosity; just under half have luminosities above $10\times$ the quiescent level. Accounting for systematic affects from our choice of search algorithm (see Appendix for more details), we estimate a flare frequency of $1.1_{-0.1}^{+0.2}$ flares per day above a luminosity of $10^{34}$ erg s$^{-1}$, with bright flares ($L>10^{35}$ erg s$^{-1}$) occurring every $\sim11.5$ days. Despite their rarity, the three brightest \textit{Chandra} flares represent roughly a third of the observed flare emission in 2012, which is itself one third of the total X-ray output of the inner $1.25\arcsec$ during the XVP. By our estimates, the unobserved flares make a $\lesssim10\%$ contribution to the quiescent flux from the accretion flow.

Taken at face value, the properties of the 39 observed flares provide relatively few clues to their physical origin, and seem to follow from the independence of the flare intensity on duration (which confirms what was observed in 2007 April with \textit{XMM-Newton}; \citealt{Porquet08}). Only two unusual features are immediately apparent in Figure \ref{fig:pr_fl_dt}: (1) the gap in the peak flare count rate and (2) the narrow distribution of peak rate versus fluence at high fluence. For reference, the gap-center luminosity is around $5\times10^{34}$ erg s$^{-1}$, approximately 15 times the quiescent luminosity of the inner $1.25\arcsec$ (\citealt{Nowak12}). The gap in the peak count rate is unexpected but only marginally significant, and after controlling for pileup and carefully subtracting the quiescent emission, we find no evidence for any spectral variations above and below this gap.  Thus it does not appear to provide evidence for distinct mechanisms of flare emission. 

 \citet{Degenaar12_arxiv} reported six additional X-ray flares in 821 ks of \textit{Swift} observations of the Galactic Center, all with luminosities above $10^{35}$ erg s$^{-1}.$ Due to \textit{Swift}'s wider PSF  and the presence of relatively bright transients very close to \sgra (\citealt{Muno05a}; \ CXOGC 1J74540.0--290031: \citealt{Muno05b,Porquet05}; and the newly-discovered pulsar SGR J1745--2900: \citealt{Degenaar13_ATel,Dwelly13_ATel,Mori13,Kennea13,Rea13_ATel}), some caution is warranted in attributing this variability to \sgra. However Monte Carlo simulations based on the \textit{Chandra} flare frequency and the \textit{Swift} observing strategy confirm that the flare frequencies are consistent, although the \textit{Swift} rate may be slightly higher. Further study of \textit{Chandra, Swift,} and \textit{XMM-Newton} observations is necessary to place hard limits on the flare frequency derivative. Because the \textit{Swift} observations are much shorter than the typical flare duration, it is difficult to compare any other properties directly since there is a degeneracy between mean luminosity and peak luminosity for short observations. Although the brightest \textit{Swift} flare was marginally softer, the six flares were consistent with a single X-ray spectral index (\citealt{Degenaar12_arxiv}). \citet{Porquet08} also found no evidence for a correlation between flare luminosity and X-ray photon index in a sample of bright \textit{XMM-Newton} flares, so it may be that no such correlation exists.

The shape of the peak rate vs fluence distribution at high fluence may be more telling. Because flares of similar intensity would be easily detectable at any duration, it seems that the brightest flares have a characteristic time scale of about 5 ks (again, the brightest \textit{XMM-Newton} flares are slightly shorter, about 3 ks; \citealt{Porquet03,Porquet08}). This time scale is sufficiently generic that it can be reproduced by a number of models (see \citealt{Markoff01,Liu02,Liu04,Dodds-Eden10,Kusunose11}). For example, it is comparable to the orbital period within a few times the radius of the innermost stable circular orbit, as well as the characteristic flyby time scale for asteroids at a distance of $\sim1$ AU (\citealt{Zubovas12}; see also \citealt{Cadez08,Kostic09}) and the Alfv\'{e}n crossing time for magnetic loops near the black hole (\citealt{Yuan03,Yuan04}). 

The same can be said of the flare luminosity distribution, which falls off with X-ray luminosity as $dN/dL\sim L^{-1.9}.$ This is consistent with several models. For example, the measured luminosity and peak rate distributions are similar to the distributions of solar flares (e.g.\ \citealt{Crosby93,Gudel04,Crosby11,Aschwanden12}), although it is not likely that coronal flares can reproduce the luminosity of flares reported in this paper (\citealt{Sazonov12} and references therein; the quasi-symmetric lightcurves are also unusual for stellar flares). The luminosity distribution is also consistent with the tidal disruption of asteroids (under the assumption that the mass function of asteroids in the central parsec is a power law with an index close to that obtained from collisional fragmentation cascade calculations for solid bodies in orbit around a solar-type star; \citealt{Zubovas12}). Since the true asteroid mass function in the harsh environment of \sgra is unknown, the agreement between the data and this model is only tentative, and in any case this comparison should not be taken as a rejection of other models.  Any viable model should be able to explain at least the luminosity distribution of the flares, the fluence distribution (which is dominated by the highest-fluence flares, since $F^2dN/dF\sim F^{0.5}$), as well as the lack of a strong correlation between the flare durations/intensities and the absence of $\gtrsim10$ ks flares.

\subsection{Comparison with the NIR}
\label{sec:multi}
An alternative perspective on the origin of the flares comes from a comparison of the X-ray and infrared variability properties (we leave the discussion of the radio/submillimeter flares for a later date; \citealt{Yusef-Zadeh06a,Yusef-Zadeh06b,Marrone08,Falcke09}). Based on their analysis of over 10,000 NACO/VLT $Ks$-band images from 2004--2009, \citet{Dodds-Eden11} found that the NIR flux distribution could be described by a lognormal distribution with a power law tail at high flux. They suggested that this tail was evidence of two states (quiescent and flaring, for the lognormal and power law components, respectively) in the NIR emission. For their largest sample, they found a power law slope of $-2.70\pm0.14,$ which is somewhat steeper than our observed slope of $-1.9^{+0.3}_{-0.4}.$ For a smaller subset of their data with high-quality photometry, however, they found a marginally significant tail with a slope of $-2.1\pm0.6,$ which is consistent with our observed X-ray luminosity distribution.

There are several reasons to proceed carefully with this comparison. The first is that we have estimated the fluxes and luminosities of our flares by supposing that these quantities are strictly proportional to the mean count rate of a flare, and calibrating by the brightest flare ever observed. We have no evidence at present for any color evolution with luminosity, but for flares with small numbers of photons under-sampling of the X-ray spectrum may lead to significant uncertainties in this scaling (which we will estimate in future work), and definitive results will require spectral analysis of each flare.

The second reason to proceed with caution is that the actual distribution described by \citet{Dodds-Eden11} is the distribution of all observed NIR fluxes from \sgra, while the quantity we have reported in this paper is the distribution of the average luminosities of X-ray flares. A strict comparison, which is beyond the scope of this paper but will be considered in future work, would require the distribution of the X-ray flux in each 300 s time bin during the 2012 XVP. If the similarity of the distributions remains after a proper and rigorous comparison, it would have interesting implications for the multiwavelength variability properties of \sgra. For instance, if the NIR  and X-ray flare emission mechanisms are synchrotron and SSC, respectively, then the fact that their flux distributions were similar might indicate that the Thompson optical depth of the flare-producing electrons is not a strong function of the intensity of the flares (see, e.g.\ \citealt{Marrone08,Witzel12}).

A third reason for caution comes from a later analysis (\citealt{Witzel12}) of a larger (2003--2010) NACO/VLT NIR dataset. The analysis of over 10,000 $Ks$-band images revealed that the NIR flux distribution is apparently consistent with a pure power law distribution (decreasing with near-infrared flux as $F_{\rm NIR}^{-4.215\pm0.05}).$ This conclusion is also supported by timing analysis by \citet{Meyer08} and \citet{Do09}, who argue that the power spectrum of \sgra in the NIR is featureless (although see \citealt{Meyer09} for evidence of a power law break around a time scale of $\sim150$ minutes). If these analyses are correct, then there is, in a sense, no such thing as an infrared ``flare" from \sgra: all the observed variability is a noise continuum. 

In the context of our analysis, this would imply that the luminosity of a flare is not well defined. If it can be demonstrated in subsequent analysis that the X-ray flux distribution from the 2012 \textit{Chandra} XVP is consistent with the NIR flux distribution as reported by \citet{Dodds-Eden11} and \citet{Witzel12}, then perhaps there is also no such thing as an X-ray flare. Instead, the peaked X-ray variability observed by \textit{Chandra} \textit{XMM-Newton}, and \textit{Swift} would simply be the high-energy, high-luminosity end of a continuous multiwavelength red noise process. Again, only those X-ray excursions that are sufficiently bright to appear above the ``blanket" of the quiescent thermal emission (\citealt{Markoff10}) are detectable. In any case, with the brightest X-ray flares characterized by moderate spectral indices ($\Gamma\sim2-2.4,$ \citealt{Nowak12,Porquet08,Porquet03}) that can be reproduced by both synchrotron and SSC models (e.g. \citealt{Yuan03,Dodds-Eden09,Eckart12}, and references therein), it seems that a systematic statistical comparison of the NIR/X-ray flux distributions may place the tightest constraints on the physics of variability in the Galactic Center (see also \citealt{Eckart06}). It may also be worth using the recent multiwavelength coverage to revisit the relationship of the flares to the fundamental plane (\citealt{Markoff05b,Plotkin12}). Whatever the origin of the bright X-ray emission, it is clear that future studies of X-ray variability from \sgra will benefit greatly from the improved flare statistics provided by the 2012 \textit{Chandra} XVP, with major progress to be made from focused comparisons of observations of the Galactic Center across the electromagnetic spectrum.

%It may be possible to address the physics of X-ray variability by other means as well. For example, we can study whether there is any statistically significant clustering (as suggested by \citealt{Porquet08}, who discovered four bright flares within just 40 ks). In our recent observations, we have a number of observations with multiple flares, including four moderate ($\gtrsim10\times$) flares in the course of 20 ks. A detailed study of the waiting times between flares and the relationship to the flare intensity may provide interesting insights into the relevant electron energization mechanisms; the asymmetry in the flare profiles could also probe the energy injection processes. 

\subsection{Flares and \sgra's Intrinsic X-ray Emission}
\label{sec:flarepsf}
Whether or not ``flares" are well defined, it is clear from our present analysis that the X-ray and infrared do differ in one respect: there are at least two X-ray emission mechanisms at work around \sgra. We have demonstrated here that the quiescent emission cannot be considered to be the superposition of a number of weak flares, on the grounds of both energetics and X-ray color (of course, the X-ray color analysis is also supported by prior spectral studies of individual flares, e.g.\ \citealt{Baganoff01,Goldwurm03,Porquet03,Belanger05,Porquet08,Nowak12}). Superimposed on this quiescent background is a variable component whose origin (both in terms of the processes that impart energy to the electrons and the mechanisms by which that energy is released as X-rays) is unknown. Our analyses of the flare distributions and the quiescent lightcurve suggest that the faint (unobserved) end of this variable component could contribute as much as 10\% of the apparent quiescent flux (see also \citealt{Wang13}). In addition, weak flares with a characteristic time scale of 1 ks could produce the observed $\sim10\%$ excess correlated variability in Figure \ref{fig:poisson}.

Remarkably, a point source with $\sim10\%$ of the quiescent luminosity is also required in order to fit the 1999--2010 radial surface brightness profile of \sgra (\citealt{Shcherbakov10}). Could this 10\% point source emission be completely comprised of the weak flares described above? If so, there must be very little truly quiescent emission near the event horizon: the immediate environment of \sgra, i.e.\ the inner $\sim10 R_{\rm S}$, emits almost no X-rays outside of flares. In other words, the quiescent X-ray emission from the inner $1.25\arcsec$  can be cleanly divided into weak X-ray flares from the inner accretion flow (the point source, $\lesssim10\%$) and steady thermal emission from plasma on scales comparable to the Bondi radius (the extended source, $\gtrsim90\%$). This could explain why the underlying emission exhibits little variability compared to other accreting black holes (Figure \ref{fig:lc1}; \citealt{Nowak12}), and more importantly why \sgra appears to lie on the fundamental plane during flares (though it does not explain why \sgra is not always on the fundamental plane or why only \sgra exhibits these flares).

Implicit in this explanation are the assumptions that the flare luminosity distribution (1) can reliably be extended to very low fluxes, and (2) is stable over many, many dynamical times in the inner accretion flow (see the beginning of Section \ref{sec:discuss} and \citealt{Degenaar12_arxiv}). The first point is interesting because for sufficiently faint flares, \sgra would have to be flaring constantly, and recent radiative GRMHD simulations (\citealt{Drappeau13}) produce mildly variable X-ray emission at roughly the appropriate level ($10^{32-33}$ erg s$^{-1}$). The stability of the flare luminosity distribution may also pose a challenge to flare models, but with the exception of asteroids these models do not predict flare distributions or their time dependence. %Thus it remains to be seen whether the stability of the flare luminosity distribution places any constraints on the astrophysics of flares from \sgra.

The $\lesssim10\%$ luminosity of the inner accretion flow may place an important independent constraint on theoretical models, which have historically predicted a larger contribution (especially for the high inclinations, and in many cases spins, that are preferred by millimeter VLBI observations and spectral observations, e.g.\ \citealt{Moscibrodzka09,Dexter10,Broderick11,Shcherbakov12,Drappeau13} and references therein). For instance, the ``best bet" model of \citet{Moscibrodzka09}, i.e.\ inclination $i=85^{\circ}$ and spin $a_{*}=0.94,$ predicts $L_{\rm X}=10^{32.9}$ erg s$^{-1}$, which is roughly twice the upper limit we have reported here. 

Finally, we note that because the average X-ray luminosity of the accretion flow is $\lesssim10\%$ of the observed quiescent emission, the brightest flare (\citealt{Nowak12}) was actually $\lesssim1300\times$ more luminous than its local background. This peak-to-mean ratio is much larger than in the infrared ($\sim10-30$; \citealt{Kunneriath10,Schoedel11,Dodds-Eden11}), and the variable X-ray/IR ratio may place important constraints on the flare production and emission mechanisms. For example, synchrotron models with cooling breaks (e.g.\ \citealt{Kardashev62,Dodds-Eden09,Trap10}) may lead to spectral slopes independent of flare luminosity, but may struggle to produce the weak flares discussed above. Inverse Compton scenarios, on the other hand may naturally lead to larger flare amplitudes in the X-ray than NIR, but may also involve changes in spectral slope. Given the progress and promise of X-ray and multiwavelength studies of flares, as well as detailed observational and theoretical analysis of the quiescent X-ray spectrum and variability, there is significance cause for optimism that our 2012 campaign will reveal the nature of the accretion flow onto the supermassive black hole.

\begin{figure*}
\centerline{\includegraphics[width=\textwidth]{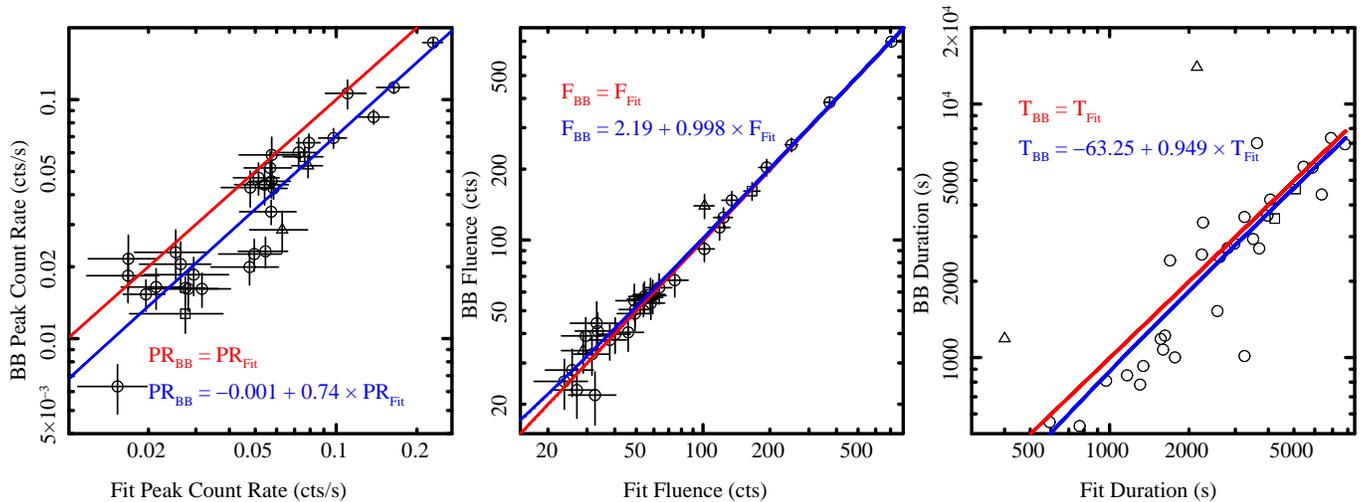}}
\caption{A comparison of the measured properties of the 34 flares detected by both the Bayesian Blocks algorithm and by direct fitting. (Left): There is a strong correlation between the two measures of flare peak count rate, although the direct fitting method returns larger rates (this is likely because the Bayesian Blocks method treats flares as uniform-rate bins). (Middle): The fluences measured by the two methods are statistically indistinguishable. (Right): Flare durations are strongly correlated, with more scatter about the mean relation. Note: circles, squares, and triangles identify flares that are truncated by the beginning/end of an observation according to neither method, both methods, and the Bayesian Blocks method, respectively.}
\label{fig:appendix}
\end{figure*}

\acknowledgements We thank all the members of the \sgra \textit{Chandra} XVP collaboration\footnote{http://www.sgra-star.com/collaboration-members}, and we are immensely grateful to \textit{Chandra} Mission Planning for their support during our 2012 campaign. We thank John Houck, John Davis, and Feng Yuan for very useful discussions concerning the analysis and interpretation of our data. We acknowledge the role of the Lorentz Center, Leiden, and the Netherlands Organization for Scientific Research Vidi Fellowship 639.042.711 (S.M.). J.N.\ gratefully acknowledges funding support from NASA through the Einstein Postdoctoral Fellowship, grant PF2-130097, awarded by the \textit{Chandra} X-ray Center, which is operated by the Smithsonian Astrophysical Observatory for NASA under contract NAS8-03060, and from NASA through the Smithsonian Astrophysical Observatory contract SV3-73016 to MIT for support of the \textit{Chandra} X-ray Center, which is operated by the Smithsonian Astrophysical Observatory for and on behalf of NASA under contract NAS8-03060. F.K.B.\ acknowledges support from the \textit{Chandra} grant G02-13110A under contract NAS8-03060. N.D.\ is supported by NASA through Hubble Postdoctoral Fellowship grant number HST-HF-51287.01-A. R.W.\ was partly supported by an European Research Council starting grant.

\begin{appendix}%section*{APPENDIX}
\label{sec:app}
Given the uncertainties associated with searching noisy signals for flares, it is worth attempting to evaluate the robustness of our flare detection algorithm and the subsequent conclusions. We proceed in several ways. First, we try slightly different binning schemes for our lightcurves (shifting the bins by one half width, performing the search on barycenter-corrected data). The bin shifts produce essentially the same flare properties, with small changes ($n\sim1$) in the number of flares depending on the specific binning. We have not explored different bin widths. Our Monte Carlo simulations (Section \ref{sec:dist}) provide an additional sense of the robustness of our algorithm, insofar as the false detection probability is negligible for our observed flares.

A more rigorous test of our systematic uncertainties can be made by using a completely different flare detection algorithm. For this, as noted in Section \ref{sec:obs}, we employ a Bayesian Blocks algorithm (\citealt{Scargle12_arxiv}). This method operates on the individual unbinned events and identifies statistically-significant changes in the count rate (see, e.g.\ \citealt{Baganoff03,Nowak12}). For each change, we require a detection significance of 96.9\%, i.e.\ $1-\exp(-3.5),$ which implies an overall significance for each detected flare of at least $\sim99.9\%.$ The method is explained thoroughly in \citet{Nowak12}, but the result is the decomposition of the X-ray lightcurve into a sequence of blocks and count rates, from which the flare properties can be measured.

The Bayesian Blocks algorithm returns a set of 45 flares. Of the features identified as flares, 34 are detected by both algorithms, with 5 detected only by the direct fitting method (Section \ref{sec:lc}), and 9 detected by the Bayesian Blocks algorithm alone. A direct comparison of the 34 definitively detected flares (Figure \ref{fig:appendix}) reveals that both algorithms return similar properties for the flares. In the left panel, we show the measured peak rates with best fits. There is a strong correlation between the peak rates as estimated by the two methods, although the Bayesian Blocks finds values that are $\sim75\%$ smaller. This is likely due to the decomposition of smooth flares into uniform rate bins. The right panel shows the measure durations, which are in general slightly shorter according to the Bayesian Blocks method. In the middle panel, we compare the measured fluences. If the fluence measured by the Bayesian Blocks routine and direct fit method are $F_{\rm BB}$ and $F_{\rm Fit},$ we find that $F_{\rm BB}=2^{+9}_{-8}+1.00^{+0.09}_{-0.10}~F_{\rm Fit}$. In other words, both methods return essentially identical fluences. This is excellent assurance that despite a few small differences in the fit parameters, our characterization of the observed X-ray events is robust. 

%2.1936e+00    8.8782e+00    7.5214e+00
%9.9760e-01    8.5072e-02    9.6105e-02

In addition to providing a measure of the uncertainty in the number of flares present in the XVP lightcurves and the reliability of our measurements, the comparison of the results of our two independent methods also illustrates the strengths and weaknesses of our flare detection
algorithms. For instance, much of the discrepancy in the number of detected flares can be accounted for by the sensitivity of the Bayesian Blocks routine to long faint flares, which may be statistically significant but not particularly Gaussian. In fact, most of the 9 flares not detected by the direct fitting method have peak count rates under 0.01 counts per second, which is below the sensitivity limit of the direct fitting method (Section \ref{sec:dist}). On the other hand, we note that the Bayesian Blocks routine appears to underestimate the peak count rates of the flares (relative to the direct fitting method), an effect that is likely due to treating the flares as blocks rather than as smooth curves.

\end{appendix}
\bibliographystyle{apj_set3}
\bibliography{ms}

\label{lastpage}

\end{document}